\documentclass[aps,prb,onecolumn, footinbib, 10pt]{revtex4-2}

\usepackage{amsmath,amssymb,bm} 
\usepackage{graphicx,comment}

\usepackage[tight]{subfigure} 

\usepackage[dvipsnames]{xcolor} 
\usepackage[papersize={8.5in,11in}]{geometry}
\usepackage[colorlinks=true]{hyperref}
\hypersetup{
    bookmarks=true,         
    unicode=false,          
    pdftoolbar=true,        
    pdfmenubar=true,        
    pdffitwindow=false,     
    pdfstartview={FitH},    
    pdfkeywords={keyword1} {key2} {key3}, 
    pdfnewwindow=true,      
    colorlinks=true,       
    linkcolor=magenta, 
    citecolor=blue,        
    filecolor=magenta,      
    urlcolor=blue           
} 

\usepackage{graphicx}
\usepackage{dcolumn}
\usepackage{color}
\usepackage{amssymb,amsmath, makecell}
\usepackage{tabularx, graphicx}
\usepackage{epstopdf}
\usepackage{latexsym}
\usepackage{colortbl}
\usepackage{psfrag}
\usepackage{bbm,bm,array,physics}
\usepackage{dsfont}
\usepackage{float, mathrsfs}

\def \nn{\nonumber \\}
\def\*#1{\mathbf{#1}} 

\begin{document}
\title{Direction-dependent linear response for gapped nodal-line semimetals in planar-Hall configurations}

\author{Fasil Hussain Rather}
\author{Firdous Haidar}
\author{Muhammed Jaffar A.}
\author{Ipsita Mandal}
\email{ipsita.mandal@snu.edu.in}

\affiliation{Department of Physics, Shiv Nadar Institution of Eminence (SNIoE), Gautam Buddha Nagar, Uttar Pradesh 201314, India}

\begin{abstract}
We compute the magnetoelectric conductivity for \textit{ideal} nodal-line semimetals (NLSMs), with a finite but tiny mass-gap, in distinct planar-Hall set-ups. Each differing configuration results from the relative orientation of the nodal-ring's plane with respect to the plane spanned by the electric ($\mathbf E $) and magnetic ($\mathbf B$) fields. The net conductivity tensor has components comprising the Drude, anomalous-Hall, in-plane (with $\mathbf E$ and $\mathbf B$) longitudinal and transverse, and Lorentz-force-operator-induced parts. Our results feature the signatures of the inherent topology of a gapped NLSM, revealed through nonzero values of the Berry curvature and the orbital magnetic moment. In particular, we show that both of these vector fields, arising in the momentum space, give rise to terms of comparable magnitudes in the resulting response. Our explicit theoretical expressions will help identify unique signatures of NLSMs in contemporary experiments.
\end{abstract}

\maketitle

\tableofcontents


\section{Introduction}

\begin{figure*}[t]
\subfigure[]{\includegraphics[width=0.25 \textwidth]{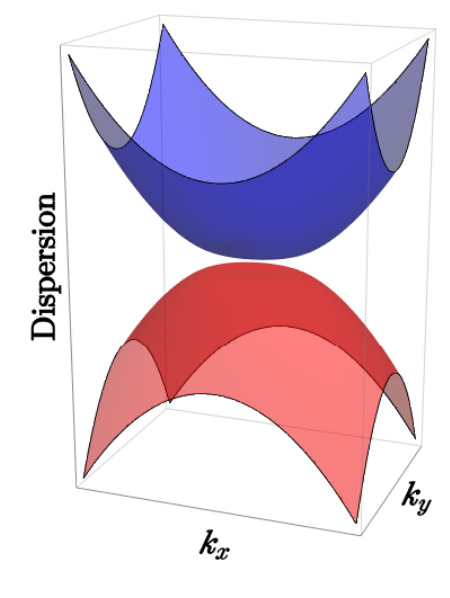}} \hspace{ 1 cm}
\subfigure[]{\includegraphics[width=0.3 \textwidth]{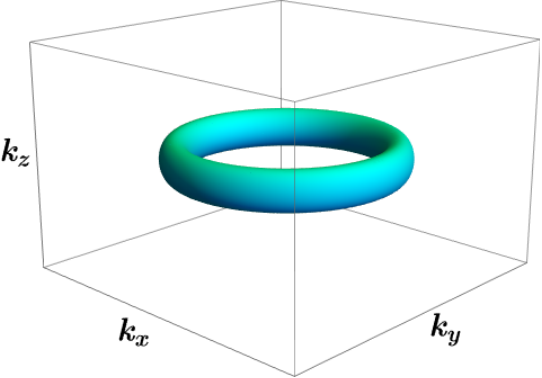}  
\includegraphics[width=0.3 \textwidth]{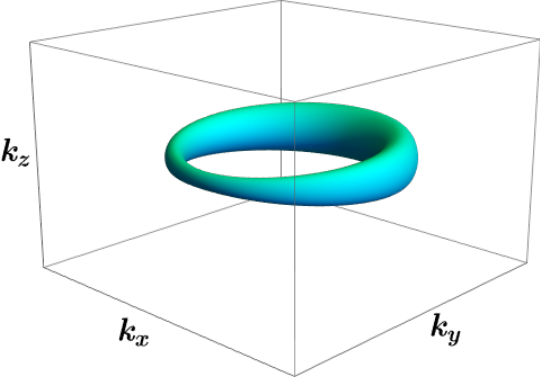}}
\caption{\label{figfs}Gapped nodal-line semimetal with isotropy along the $k_x k_y$-plane:
(a) Dispersion against the $k_x k_y $-plane.
(b) Schematics of the Fermi surfaces representing the scenarios for without and with the OMM-correction, respectively. Here, we have taken the applied magnetic field ($\mathbf B$) to be directed purely along the $y$-axis.  A toroid-shaped Fermi surface deforms into a ring cyclide when a nonzero $\mathbf B $ is applied. We have assumed that $ |\mathbf B| $ is low-enough so as not to cause a Lifshitz transition of the Fermi surface to a horn cyclide.}
\end{figure*}

The discovery of three-dimensional (3d) semimetals, featuring symmetry-protected band-crossings, have brought about a direct application of the mathematical concepts of topology into understanding the bandstructures of materials. They exemplify materials whose Brillouin zones (BZs), when treated as closed manifolds, are endowed with nontrivial topological properties. Such a band-crossing can occur at a nodal point~\cite{burkov11_Weyl, yan17_topological, bernevig, grushin-multifold} or a nodal line~\cite{balents-nodal}, thus forming a zero-dimensional or a one-dimensional (1d) Fermi surface, respectively, when the chemical potential is adjusted to cut the band-crossing energy. Hence, they represent singular points of the BZ (spanned by the momentum coordinates of $\mathbf k \equiv \lbrace k_x, k_y, k_z \rbrace $) where the density-of-states goes to zero. In contrast with the nodal points serving as sources/sinks of the Berry curvature (BC), the nodal-line semimetals (NLSMs) exhibit a quantized Zak phase \cite{schnyder_nodal, zak_nodal, fu_nlsm}. For example, in a $\mathcal P \mathcal T$-symmetric \footnote{$\mathcal P$ and $\mathcal T$ represent the inversion and time-reversal symmetries, respectively.} two-band NLSM, a loop encircling the nodal line accumulates a Berry phase equalling an integer times $\pi$ \cite{schnyder_nodal, biao_nodal, zak_nodal}. The BC vanishes in the entire BZ, except at the nodal line, where it becomes singular, thus reﬂecting the topological nature of the NLSMs. 
While surface states in the form of 1d Fermi arcs constitute fingerprints of 3d nodal points (residing in the bulk of the BZ), nodal lines in the bulk of BZs reveal themselves via the so-called drumhead surface-states \cite{balents-nodal}, which can be observed using high-resolution angle-resolved photoemission spectroscopy (ARPES) \cite{arpes-nlsm}.
On introducing a small $\mathcal P \mathcal T$-symmetry-breaking mass-term ($\propto \Delta $), the nodal line is gapped out, and the entire BZ acquires a well-behaved nonvanishing BC. Thus, a finite $\Delta$ changes a 1d nodal-ring Fermi surface to a 2d toroidal manifold that encircles the nodal loop. Here, we will consider the case when the circle, defining the major axis of the torus, lies perpendicular to the $k_z$-component of the momentum vector and possesses a rotational symmetry about the $k_z$-axis (cf. Fig.~\ref{figfs}). This results in a gapped nodal-ring with the BC-flux lines forming vortices around the $k_z$-axis. 

The Berry phase is the fundamental quantity which causes topological properties like the BC to appear in the space spanned by the BZ \cite{xiao_review, sundaram99_wavepacket,graf-Nband, graf_thesis, timm, ips_rahul_ph_strain, rahul-jpcm,  ips-kush-review, claudia-multifold, ips-ruiz, ips-tilted, ips-rsw-ph, ips-shreya, ips-spin1-ph}. In addition to the BC, the Berry phase sources another vector field called the orbital magnetic moment (OMM), which shows up when a semimetal is subjected to a nonzero magnetic field, as a consequence of the semiclassical self-rotation of the quasiparticle wavepacket \cite{xiao_review, sundaram99_wavepacket}. Examples of some transport-measurements, where the BC and OMM affect the resulting signatures, encompass the intrinsic anomalous-Hall effect~\cite{haldane04_berry,goswami13_axionic, burkov14_anomolous}, planar-Hall conductivity \cite{zhang16_linear, chen16_thermoelectric, nandy_2017_chiral, nandy18_Berry, amit_magneto, das20_thermal, das22_nonlinear, pal22a_berry, pal22b_berry, fu22_thermoelectric, araki20_magnetic, mizuta14_contribution, ips-mwsm-floquet, ips_rahul_ph_strain, timm, rahul-jpcm, ips-kush-review, claudia-multifold, ips-ruiz, ips-tilted, phe_nlsm, ips-rsw-ph, ips-shreya, ips-spin1-ph}, magneto-optical conductivity under strong magnetic fields~\cite{gusynin06_magneto, staalhammar20_magneto, yadav23_magneto}, Magnus Hall effect~\cite{papaj_magnus, amit-magnus, ips-magnus}, circular dichroism \cite{ips-cd1, ips_cd}, circular photogalvanic effect \cite{moore18_optical, guo23_light, kozii, ips_cpge}, and quasiparticle-tunneling across potential barriers/wells \cite{ips-aritra, ips-jns, ips-sandip, ips-sandip-sajid}. 
Just like the topological properties of 3d nodal-line semimetals leave their trademark signatures in various transport-properties, the gapped NLSMs give rise to novel features in the Berry-phase-induced response-coefficients \cite{schnyder_nodal, yang1, yang_review_nlsm, chen_nlsm, ips-magnus, flores, phe_nlsm, claudia_nlsm, enke}. In particular, since an NLSM can contribute to a significant BC-flux over a substantial volume of the BZ, it enhances the magnitude of the anomalous-Hall effect \cite{enke}.

NLSMs have been reported to exist in a variety of distinct materials, such as SrAs$_3$ \cite{arpes-nlsm}, Ca$_3$P$_2$ \cite{expt1_nlsm}, hexagonal pnictides (CaAgP and CaAgAs) \cite{expt2_nlsm}, photonic metamaterials \cite{biao_nodal}, alkaline-earth metals (e.g., Ca, Sr and Yb)~\cite{alkaline_nlsm}, Fe$_2$MnX \cite{claudia_nlsm}, and Co$_3$Sn$_2$S$_2$ \cite{enke}. Based on \textit{ab initio} simulations, CuTeO$_3$ \cite{cuteo_nlsm} is predicted to host an ideal NLSM, which implies that the nodal loop is close to the Fermi level, relatively flat in energy (e.g., lying along the $k_x k_y$-plane), simple in its shape (e.g., can be assumed to be circular), and not coexisting with other extraneous bands. Additionally, consideration of a nonzero spin-orbit-coupling (SOC) is shown to open up only a tiny gap. This system thus exemplifies the model Hamiltonian that we are going to consider here, validating our idealization of a nodal line shown in Fig.~\ref{figfs}. 

In this paper, our focus is on the analytical computation of the linear response in the form of magnetoelectric conductivity, when we subject an ideal NLSM to the combined action of static and uniform electric ($\mathbf E $) and magnetic ($\mathbf B$) fields. This constitutes a planar-Hall set-up, where $\mathbf B $ is generically applied at a non-perpendicular angle ($\theta$) with respect to $\mathbf E$ --- this ensures that the projection of $\mathbf B$ along the axis of $\mathbf E$ is nonzero, and the two fields define a \textit{plane}. The presence of the nodal line allows us to play around with the orientation of the $\mathbf E \, \mathbf B$-plane with respect to the nodal-line-plane, thus opening up the possibility of anisotropic transport. Analogous situations have been studied in the context of multi-Weyl semimetals, utilizing the anisotropy in their dispersion \cite{rahul-jpcm, ips-tilted}. Here, we study three distinct set-ups as shown in Fig.~\ref{figsetup}(b).

The paper is organized as follows. In Sec.~\ref{secmodel}, we discuss the effective continuum model for an ideal NLSM with a small gap. Sec.~\ref{secsigma} is devoted to the computation of the magnetoelectric conductivity. Finally, we wrap up in Sec.~\ref{secsum} with a summary and some future-outlook. In all expressions that follow, we resort to using the natural units --- this means that the reduced Planck's constant ($\hbar $), the speed of light ($c$), and the Boltzmann constant ($k_B $) are each set to unity.
The magnitude of electric charge, $e$, has no units and also equals unity in the natural units. Nevertheless, merely for the sake of book-keeping, we will retain $e$ in our expressions.

\section{Model}
\label{secmodel}

The minimal model of an NLSM, comprising two bands and a single circular nodal loop lying in the $k_x k_y$-plane, is captured by \cite{balents-nodal, yang1}
\begin{align}
\mathcal{H}_0 (\mathbf k ) = {\mathbf d}_0 (\mathbf k) \cdot \boldsymbol{\sigma} \,,
\quad {\mathbf d}_0 (\mathbf k)
= \left \lbrace  \lambda
\left( k_\perp^2-k_0^2 \right), \, v_z\, k_z , \, \Delta \right \rbrace, \quad
k_\perp = \sqrt{k_x^2 + k_y^2 }\,,
\end{align}
where $ \boldsymbol{\sigma} = \left\lbrace \sigma_x,\sigma_y,\sigma_z\right \rbrace $ is the vector comprising the three Pauli matrices as its three components. Here, $\lambda $ and $k_0$ are material-dependent parameters, and $\Delta$ represents the tiny gap opened up by symmetry-breaking (for example, by SOC).
For $\Delta = 0$, the two bands cross at $k_\perp^2-k_0^2 = 0 $ and $k_z = 0$, defining a nodal ring of radius $ k_0 $. For a chemical potential ($\mu$) that satisfies $\mu \ll \lambda \, k_0^2$, we have low-energy excitations confined in the vicinity of the resulting Fermi surface (encircling the nodal ring). Hence, for characterizing the transport-signatures of low-energy quasiparticles, it is advantageous (for computational purposes) to linearize $\mathcal{H}$ in the momentum deviation from the location of the nodal line \cite{linearize-nlsm}. This is accomplished by implementing a transformation to the toroidal coordinates as follows:
\begin{align}
\label{eqtrs}
k_x = \left( k_0+ \kappa \cos{\phi} \right) \cos \Phi \,, \quad
k_y = \left ( k_0 +  \kappa \cos{\phi} \right ) \sin \Phi \,,\quad
k_z = \frac{ \kappa\sin{\phi}}{\alpha}\,, \quad \alpha = \frac{ v_z}  {v_0} \,, 
\quad v_0 = 2 \, \lambda \,k_0\,.
\end{align}
The Jacobian of the coordinate transformations is $ J =   \kappa\left(\,k_0 +  \kappa\cos{\phi}\, \right) / \alpha $.
Inverting the transformation relations, we have $ k_0 + \kappa \cos \phi = \pm \, k_\perp $. But since $\kappa \ll k_0 $ in the low-energy limit, we have $ \kappa \cos \phi  = k_\perp - k_0 $.
Hence, the truncated expressions take the forms of
\begin{align}
& \mathcal{H}_0 (\mathbf k ) = 
\mathcal{H} (\delta \mathbf k) + \order{\kappa^2}\,, \quad 
\mathcal{H} (\delta \mathbf k) = 
{\mathbf d} ( \delta \mathbf k) \cdot \boldsymbol{\sigma} \,,
\quad
\delta \mathbf k =
\kappa  \left \lbrace \cos \phi \cos \Phi, \,
\cos \phi \sin \Phi, \,
\frac{\sin \phi} {\alpha} 
 \right \rbrace, \nn 
& \mathbf{d}( \delta \mathbf{k}) 
= \left \lbrace v_0 \, \kappa \cos \phi ,
\, v_0 \, \kappa \sin \phi,\, \Delta \right\rbrace
=
\left \lbrace v_0  \left( k_\perp - k_0 \right) ,
\, v_z \, k_z ,\, \Delta \right\rbrace , \quad
J =  \frac{ \kappa \, k_\perp } { \alpha} \,.
\end{align}
In terms of the toroidal coordinates, while $k_0$ represents the major radius (i.e., the distance between a point on the nodal ring and the center of the torus), $\kappa $ denotes the minor radius (i.e., the radius of the cross-section of the torus). $\Phi$ and $\phi $ are the angular coordinates $ \in [0, 2 \pi)$, representing rotation around the torus's axis of revolution and rotation around a point on the nodal-ring, respectively. The parameter $\alpha $ stands for the ratio between the velocities along the $z$-axis and
along the $xy$-plane, respectively.

\begin{figure}[t!]
\centering 
\subfigure[]{\includegraphics[width=0.24 \textwidth]{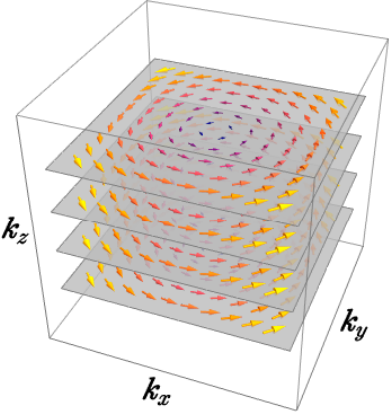}} \hspace{0.75 cm}
\subfigure[]{\includegraphics[width=0.24 \textwidth]{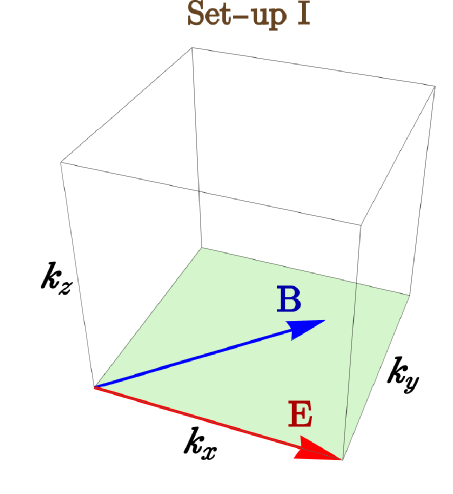}\hspace{-0.25 cm}
\includegraphics[width=0.24\textwidth]{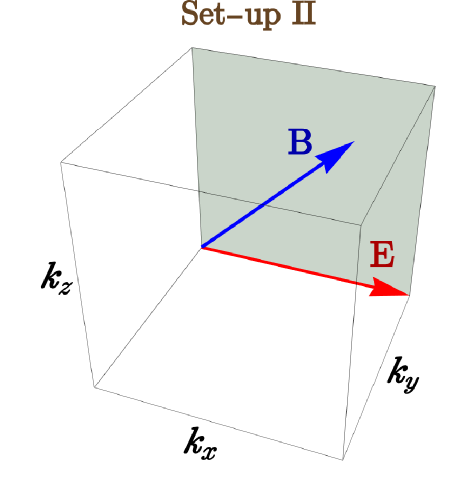}\hspace{ -0.25 cm}
\includegraphics[width=0.24 \textwidth]{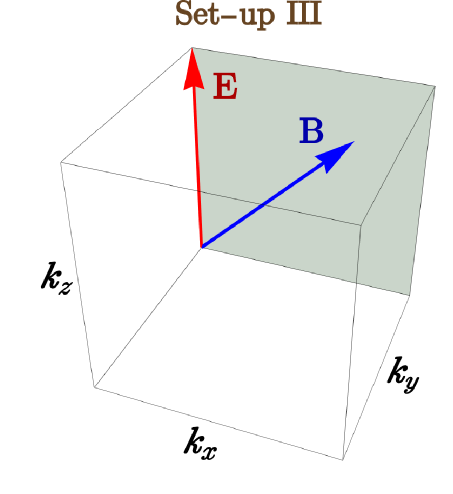}} 
\caption{(a) Vector-plot of ${\mathbf \Omega}_{s= 2} ( \mathbf k) $ in the Brillouin zone. (b) Schematics of the three set-ups that we use to investigate the planar-Hall effect in NLSMs, showing the relative alignments of the external uniform electric (red arrow) and magnetic (blue arrow) fields. We label the three scenarios as set-up I, set-up II, and set-up III, respectively. The plane containing the $\mathbf E $ and $\mathbf B $ vectors (making an angle $\theta$ with each other) in each set-up has been highlighted by a background colour-shading. The coordinates have been chosen such that the NLSM in question has its nodal line lying along the $k_x k_y$-plane (cf. Fig.~\ref{figfs}).
\label{figsetup}}
\end{figure}

Working with the linearized Hamiltonian $\mathcal H$, the eigenvalues of the two bands are obtained as
\begin{align}
\label{6}
\varepsilon_s  ({ \mathbf k}) &=(-1)^s\, \epsilon, \quad 
 \epsilon = \sqrt{ v_0^2 \,\kappa^2 +  \Delta^2}, 
\quad
s \in \lbrace 1,2 \rbrace,
\end{align}
where the values $1$ and $2$ for $s$ represent the valence (i.e., negative-energy) and the conduction (i.e., positive-energy) bands, respectively. The band velocity of the quasiparticles is given by
\begin{align}
\label{eqv}
{\boldsymbol v}^{(0,s)} ( \mathbf{k}) 
\equiv \nabla_{\mathbf{k}} \varepsilon_s  (\mathbf{k}) = 
\frac{(-1)^s \, v_0^2}{\epsilon} 
\left \lbrace k_x \left( 1- \frac{k_0}{k_\perp } \right ), 
\, k_y \left( 1- \frac{k_0} {k_\perp} \right), 
\, \frac{ v_z^2 \, k_z} {v_0^2 }  \right\rbrace.
\end{align}
The Berry curvature (BC) and the orbital magnetic moment (OMM), associated with the $s^{\rm{th}}$ band, can be evaluated using the generic formulas of
\begin{align} 
\label{eqomm}
& {\mathbf \Omega}_s ( \mathbf k)  = 
    i \, \langle  \nabla_{ \mathbf k}  \psi_s ({ \mathbf k})| \, 
    \cross  \, | \nabla_{ \mathbf k}  \psi_s ({ \mathbf k})\rangle
   \Rightarrow
\Omega^i_s ( \mathbf k) 
\overset{\text{two}-} {\underset{\text{band}} {=}}
 \frac{  (-1)^{s+1} \,  
\epsilon^i_{\,\,\,jl}}
 {4\,| \mathbf d (\delta \mathbf k) |^3} \, 
 \mathbf d (\delta \mathbf k) \cdot
 \left[   \partial_{k_j} \mathbf d (\delta \mathbf k) \cross  
 \partial_{k_l } \mathbf d (\delta \mathbf k) \right ] \text{ and }
\nn &  
{\boldsymbol{m}}_s  ( \mathbf k) 
=    \frac{ -\, i \, e} {2 } \,
\langle  \mathbf \nabla_{ \mathbf k} \psi_s({ \mathbf k})| \cross
\left [\,
\left \lbrace \mathcal{H}({ \mathbf k}) -\mathcal{E}_s 
({ \mathbf k}) 
\right \rbrace
| \mathbf \nabla_{ \mathbf k} \psi_s({ \mathbf k}) \rangle \right ]
\Rightarrow
m^i_{s} ( \mathbf k) 
\overset{\text{two}-} {\underset{\text{band}} {=}}
  \frac{- \,e \, \epsilon^i_{\,\,\,jl}
  } 
{4 \, | \mathbf d ( \delta \mathbf k) |^2} \,  
\mathbf d (\delta \mathbf k) \cdot
 \left[   \partial_{k_j} \mathbf d (\delta \mathbf k) \cross 
  \partial_{k_l} \mathbf d (\delta \mathbf k) \right ],
\end{align}
respectively. The symbol $ |  \psi_s ({ \mathbf k}) \rangle $ denotes the normalized eigenvector corresponding to the band labeled by $s$, with $ \lbrace |  \psi_1 \rangle, \,  \lbrace |  \psi_2 \rangle \rbrace $ forming an orthonormal set.
For two-band models, which are essentially of the generic form given by $\mathbf d \cdot \boldsymbol \sigma$, the relation of $ m^i_s  ( \mathbf k)  = e\, \varepsilon_s (\mathbf k) \, \Omega^i_s ( \mathbf k)$ is satisfied \cite{graf-Nband, graf_thesis}.
The indices $i$, $j$, and $l$ $ \in \lbrace x, y, z \rbrace $, and are used here to denote the Cartesian components of the 3d vectors and tensors. On evaluating the expressions in Eq.~\eqref{eqomm} for $\mathcal{H} (\delta \mathbf k) $, we get
\begin{align}
\label{eqbcomm}
\mathbf \Omega_s ({ \mathbf k})= 
\frac{(-1)^{s+1} \,v_z\, v_0\, \Delta} 
{2\, \epsilon^3 \, k_\perp }
\left\lbrace k_y \, ,-\, k_x,\,0 \right\rbrace, \quad 
   {\boldsymbol{m}}_s ({ \mathbf k}) 
= \frac{ - \,e\, v_z \, v_0\,\Delta} 
{2\, \epsilon^2 \, k_\perp } 
\left\lbrace k_y, \,-\, k_x,\,0 \right \rbrace.
\end{align}
Clearly, $\mathbf \Omega_s $ has no $ {\mathbf{\hat k_z}} $-component and has nonzero vorticity exclusively in the ${\mathbf{\hat k_x}} {\mathbf{\hat k_y}}$-plane. This is depicted in Fig.~\ref{figsetup}(a).
While the BC changes sign with $s$, the OMM does not. Hence, we will remove the subscript ``$s$'' from ${\boldsymbol{m}}_{ s }({ \mathbf k})$.

\section{Magnetoelectric conductivity}
\label{secsigma}

In this section, we will elaborate on the explicit forms of the magnetoconductivity tensors for three distinct planar-Hall set-ups, as shown in Fig.~\ref{figsetup}(b). In order to include the effects both from the BC and the OMM in the linear-response coefficients, we first define the following quantities:
\begin{align}
\label{eqtopo}
& \mathcal{E}_s  (\mathbf k) 
= \varepsilon_s   (\mathbf k) + \varepsilon^{ (m) }  (\mathbf k) \, ,
\quad 
\varepsilon^{(m)}   (\mathbf k)  = - \,{\mathbf B} \cdot {\boldsymbol{m}}  (\mathbf k) \,, \quad
{\boldsymbol  v}_s  ({\mathbf k} ) \equiv 
 \nabla_{{\mathbf k}}   \mathcal{E}_s  ({\mathbf k})
 = {\boldsymbol  v}^{(0, s)} ({\mathbf k} ) + {\boldsymbol  v}^{(m)} ({\mathbf k} ) \,,\nn
& {\boldsymbol  v}^{(m)} ({\mathbf k} )
= \nabla_{{\mathbf k}} \varepsilon^{(m)}   (\mathbf k) \,,
\quad  D_s   ({\mathbf k}) = \left [1 + e \,  \left \lbrace 
{\mathbf B} \cdot \mathbf{\Omega }_s   (\mathbf k)\right \rbrace  \right ]^{-1} .
\end{align}
Here, $ \varepsilon^{ (m) }$ is the Zeeman-like correction to the energy induced by the OMM~\cite{xiao_review, sundaram99_wavepacket, arovas, graf_thesis}, $ {\boldsymbol v}_s   $ is the modified band-velocity of the quasiparticles [after including $ \varepsilon^{ (m) } $], and $ D_s  $ is the modification factor of the phase-space volume element due to a nonzero BC \cite{arovas, ips-kush-review}. Since the OMM modifies the bare dispersion ($\varepsilon_s  $) to $ \mathcal{E}_s $, the shape of the Fermi surface is modified accordingly. This is shown schematically in Fig.~\ref{figfs} for the case when $\mathbf B $ lies in the nodal-line plane, where the original toroidal Fermi surface gets deformed into a ring cyclide. In particular, if $ B_{pl} \equiv \sqrt{B_x^2 + B_y^2} $ is increased to a critical value ($B_c$), a topological Lifshitz transition occurs with the Fermi surface transiting into a horn cyclide, pinching off at a point \cite{yang1, yang_review_nlsm}. We will assume that $  B_{pl} $ is much below $B_c$ so that we have only a slight deviation from the toroidal Fermi surface. A rough estimate of the critical value of $B_c$ can be obtained from the
relation \cite{yang1}
$$ \left[ \pm \epsilon + 
\frac{e\,v_z\, v_0\, \Delta\, B_c }{2\, \epsilon^2}\right ] \Big \vert_{\kappa = 0} 
\simeq  \mu 
\Rightarrow   B_c \simeq  \frac{ 2\, \Delta \,|\mu -\Delta|}  {e\,v_z\, v_0}\,. $$
Furthermore, we must take $ | \mathbf B |  \ll \mu^2 / (e\, v_0^2)$ in order to ensure that it is legitimate to ignore the formation of quantized Landau levels, such that their inter-level spacings are negligible. This constraint is equivalent to demanding $\kappa_F \, \ell_B \gg 1 $, where $\kappa_F \equiv \mu /v_0 $ is the Fermi momentum and $\ell_B \equiv 1/\sqrt{e\, |\mathbf B|}$ is the magnetic length (in the context of the quantum Hall effect).

The weak-magnetic-field limit implies that 
$ e \, |{\mathbf B} \cdot \mathbf{\Omega }_s  | \ll 1 . $
In our calculations, we will retain terms upto $\order{ |{\mathbf B}|^2}$. This implies that we will use the expansion of
\begin{align}
D_s  &=
 1 - e \,  \left( {\mathbf B}  \cdot \mathbf{\Omega }_s   \right) 
+   e^2  \,  \left( {\mathbf B} \cdot \mathbf{\Omega }_s   \right)^2  
+  \order{ |{\mathbf B}|^3} \,. 
 \label{Exp_D}
\end{align}
Also, the small-$|\mathbf B | $ limit ensures that $ |\varepsilon^{ (m) } (\mathbf k) | \ll |\varepsilon_s  (\mathbf k) |$, because
\begin{align}
\vert {\mathbf B} \cdot  \boldsymbol{ m } \vert \equiv
 e\, |\varepsilon_s | \, 
 \left| {\mathbf B} \cdot  \mathbf{\Omega}_s  \right |
 \ll  |\varepsilon_s  |  \,.
\end{align} 
This allows to expand the derivative of the Fermi-Dirac distribution, $f_0 ( \mathcal{E}_s,\mu, T   ) \equiv  
\left ( 1+ e^{\frac{\mathcal{E}_s -\mu} {T}} \right )^{-1} $, in a Taylor-series, where $\mu $ is the applied chemical potential and $T$ is the temperature. Retaining terms upto quadratic order in $| \mathbf B|$, we obtain
\begin{align}
& f_{\rm prime} ( \mathcal{E}_s ) \equiv 
\frac{ \partial f_0 ( \mathcal{E}_s )} {\partial {\mathcal{E}_s}} 
= f_0^\prime ( \varepsilon_s  ) 
+  \varepsilon^{ (m) } \,  
f^{\prime \prime}_0 (\varepsilon_s ) 
+ \frac{1}{2} \left(  \varepsilon^{(m)} \right)^2  \,
  f^{\prime \prime \prime}_0 ( \varepsilon_s  )  +  \order{ |\mathbf B|^3} \,,
\label{Exp_f}
\end{align}
where we have suppressed the $\mu$- and $T$-dependence for uncluttering the notations. With that understanding, a prime indicates a derivative of $f_0(u)$ with respect to $u$.

We use the expressions of the electric conductivity ($\sigma $) obtained via the semiclassical-Boltzmann formalism, applicable for a weak-magnetic-field strength, and simplified by a momentum-independent relaxation time ($\tau$). Basically, we adopt the relaxation-time approximation, which boils down to using a phenomenological scattering rate $\sim 1/\tau $. For the detailed steps, we refer the reader to our earlier works \cite{ips-kush-review, ips_rahul_ph_strain, rahul-jpcm, ips-ruiz, ips-rsw-ph, ips-spin1-ph}.
For a given alignment of the electromagnetic fields, we define the in-planar (or planar) components of $\sigma $ to be the ones which lie in the plane spanned by $\mathbf E $ and $\mathbf B $. It comprises the longitudinal (with respect to the direction of $\mathbf E$) and the in-plane transverse components, and are commonly referred to as the longitudinal magnetoconductivity (LMC) and the planar-Hall conductivity (PHC), respectively. The out-of-plane components are captured by the so-called anomalous-Hall part (denoted by $\sigma^{\text{AH}, s}$) and the Lorentz-force-operator contributions \cite{ips-rsw-ph, ips_tilted_dirac, ips-spin1-ph} (denoted by $\sigma_s^{\text{LF}}$). Actually, $\sigma_s^{\text{LF}}$ gives rise to longitudinal and in-plane transverse components as well \cite{ips_tilted_dirac, ips-spin1-ph}. We discuss their explicit forms below:
\begin{enumerate}

\item The generic expression for the in-plane components of the magnetoelectric conductivity tensor, contributed by the band with index $s$, is given by 
\begin{align}
\bar \sigma^s_{i j} &= -  \,  \tau  \, e^2 
\int \frac{  d^3 \mathbf k } { (2\, \pi )^3  } \,  
D_s   \,  \left[ \left( v_s  \right)_i + e  \, 
 (  {\boldsymbol  v}_s  \cdot \mathbf{\Omega }_s  ) \,  B_i \right] 
 \left[ \left( v_s  \right)_j + 
 e  \,  (  {\boldsymbol  v}_s  \cdot \mathbf{\Omega }_s  ) \,  B_j \right] 
\frac{\partial f_0 (\mathcal{E}_s ) }
{ \partial \mathcal{E}_s   }  .
\label{eq_elec}
\end{align}
For the ease of calculations, we decompose it as $ \bar \sigma^s_{i j} = \sigma_{ i j}^{(s,1) }
+ \sigma_{ i j}^{(s,2) } + \sigma_{ i j}^{(s,3) } + \sigma_{ i j}^{(s,4) }$, where
\begin{align}  
\label{eq_sig_4parts}
 & \sigma_{ i j}^{ (s,1) } = \tau \, e^2 
  \int \frac{d^3   {\mathbf k}} {(2  \, \pi)^{3}} \,I_{1ij} \,,\quad
\sigma_{ij} ^{(s,2)}   =
 B_i \, B_j \, \tau \, e^4
\int \frac{d^3   {\mathbf k}}{(2  \, \pi)^{3}} \,I_2\,,
\nn &  \sigma_{ij} ^{(s,3)}   =  
{B_j \, \tau \, e^3}   
 \int \frac{d^3  \mathbf  k}{(2  \, \pi)^{3}} \,I_{3i}\,, 
 \quad  \sigma_{ij} ^{(s,4)}   = 
{B_i \, \tau \, e^3  }
  \int \frac{d^3   {\mathbf k}}{(2  \, \pi)^{3}} \,I_{3j} \,,
\nn & I_{1ij} = -\,  D_s   \,
   \left(  v_s   \right)_i \left(  v_s   \right)_j \,
f_0^\prime (\mathcal E_s )\,, \quad
  I_2 = - \, D_s   \left (  {\boldsymbol v}_s     \cdot  
  {\mathbf  \Omega}_s   \right )^2 
 f_0^\prime (\mathcal E_s )\,, \quad 
 I_{3i} = -\,   D_s   \, 
\left(  v_s  \right)_i  \left ( {\boldsymbol v}_s  
   \cdot   {\mathbf  \Omega}_s   \right )
  f_0^\prime (\mathcal E_s ) \,.
\end{align}
For the sake of simplicity, we will work in the $T \rightarrow 0 $ limit, such that $f_0^\prime (\mathcal E_s) 
\rightarrow -\, \delta (\mathcal E_s - \mu  )$.
We note that the results for $T>0$ can be easily obtained by using the relation given by \cite{mermin}
\begin{align}
\label{eqsigmat}
\sigma^s_{ij} (T) =  -\int_{-\infty}^\infty d\varepsilon
 \,\sigma^s_{ij} (T=0)\, 
\frac{ \partial f_0 ( \varepsilon ,\mu,T   )} 
{\partial \varepsilon}  \,.
\end{align}
We would also like to point out that, in the final integrals, we have switched from the $\kappa$-variable (which lies in the range $[0, k_0]$) to the $\epsilon $-variable, because in this way it is easier to deal with the Dirac-delta function and its derivatives (with respect to $\epsilon $).
Up to $\order{ |\mathbf B|^2}$, we find that
\begin{align}
\label{i1}
I_{1ij} & =  
\left \lbrace
 v^{(0, s)}_i \, v^{(0, s)}_j 
+  
 v^{(0, s)}_j \, v_i^{(m)} + v^{(0, s)}_i \, v_j^{(m)} 
- e \, v^{(0, s)}_i  \,v^{(0, s)}_j
 \left( \mathbf  B \cdot \mathbf \Omega_s  \right )
\right \rbrace \delta  (\varepsilon_s - \mu) \nn
 &  \quad +  \varepsilon^{(m)}
\left \lbrace 
v^{(0, s)}_i  \, v^{(0, s)}_j 
- e \, v^{(0, s)}_i \,  v^{(0, s)}_j  
\left( \mathbf  B \cdot \mathbf \Omega_s  \right )
+ v^{(0,s)}_{ j} \, v_i^{(m)} + v^{(0, s)}_i  
\, v_j^{(m)} \right \rbrace
\delta^\prime (\varepsilon_s - \mu)
 \nn & \quad 
 + \left \lbrace 
  e  \, v^{(0, s)}_i  \left( \mathbf  B \cdot \mathbf \Omega_s  \right )
  - v_i^{(m)} \right \rbrace 
\left \lbrace e \, v^{(0, s)}_j 
 \left( \mathbf  B \cdot \mathbf \Omega_s  \right ) -  v_j^{(m)} 
 \right \rbrace \,  \delta  (\varepsilon_s - \mu)
+ \frac {
 v^{(0, s)}_i \,v^{(0, s)}_j \left( \varepsilon^{(m)} \right)^2 
 \delta^{\prime \prime } (\varepsilon_s - \mu)
} 
{2} \,,\nn
I_2 & = \left( \boldsymbol v^{(0, s)} \cdot \mathbf \Omega_s  \right)^2  
\delta  (\varepsilon_s - \mu)\,,\nn
I_{3i} &= \left [  
\left( \boldsymbol v^{(0, s)} \cdot \mathbf \Omega_s  \right )
\left  \lbrace
 v_i^{(m)} +  v^{(0, s)}_i 
- e \,   v^{(0,s)}_i \left( \mathbf  B \cdot \mathbf \Omega_s  
 \right) \right \rbrace
+ v^{(0, s)}_i  \left( \boldsymbol v^{(m)} 
\cdot \mathbf \Omega_s  \right ) \right] 
\delta  (\varepsilon_s - \mu )
 \nn & \qquad 
+ v^{(0, s)}_i \,\varepsilon^{(m)} 
\left( \boldsymbol v^{(0, s)} \cdot \mathbf \Omega_s  \right )
 \delta^\prime (\varepsilon_s - \mu)\,.
\end{align}
From Eqs.~\eqref{eqv} and \eqref{eqbcomm}, we note that $\boldsymbol v^{(0, s)} \cdot \mathbf \Omega_s $ = 0. Therefore, for ideal NLSMs, $I_2 = 0$ and
\begin{align}
\label{i12}
e\, B_j \, I_{3 i} =  e\,B_j \, v^{(0, s)}_i  \left( \boldsymbol v^{(m)} 
\cdot \mathbf \Omega_s  \right ) 
\delta  (\varepsilon_s - \mu ) ,
\end{align}
which stems from the fact that the vector field $\mathbf \Omega_s $ has no $ {\mathbf{\hat k_z}} $-component and has nonzero vorticity exclusively in the ${\mathbf{\hat k_x}} {\mathbf{\hat k_y}}$-plane. On explicit evaluation for each set-up, we also find that $I_{3 i} $ is identically zero in our system --- hence, we will not consider it anymore. Since any linear-in-$B$ term is ruled out in accordance with the Onsager-Casimir reciprocity relations~\cite{onsager1, onsager2, onsager3}, as also seen from our explicit results, the parts of $I_{1ij}$ that can give nonzero contributions are captured by
\begin{align}
\label{i13}
I^{nz}_{1ij} & =   
 \left \lbrace 
  e  \, v^{(0, s)}_i  \left( \mathbf  B \cdot \mathbf \Omega_s  \right )
  - v_i^{(m)} \right \rbrace 
\left \lbrace e \, v^{(0, s)}_j 
 \left( \mathbf  B \cdot \mathbf \Omega_s  \right ) -  v_j^{(m)} 
 \right \rbrace \,  \delta  (\varepsilon_s - \mu)
\nn & \quad +  \varepsilon^{(m)}
\left \lbrace v^{(0,s)}_{ j} \, v_i^{(m)} + v^{(0, s)}_i  
\, v_j^{(m)}- e \, v^{(0, s)}_i \,  v^{(0, s)}_j  
\left( \mathbf  B \cdot \mathbf \Omega_s  \right )
 \right \rbrace
\delta^\prime (\varepsilon_s - \mu) 
\nn & \quad + \frac {v^{(0, s)}_i \,v^{(0, s)}_j \left( \varepsilon^{(m)} \right)^2 
 \delta^{\prime \prime } (\varepsilon_s - \mu)}  {2} \,.
\end{align}

\item
The anomalous-Hall part and the Lorentz-force-operator contributions \cite{ips-rsw-ph, ips_tilted_dirac, ips-spin1-ph}, on expanding up to $\order{ |\mathbf B|^3}$, are given by
\begin{align}
\label{AH}
& (\sigma^{\text{AH}}_s)_{ij}  = - \, e^2 \,\epsilon_{ijl}
\int \frac{  d^3 {\mathbf k} }  {(2 \,\pi )^{3} } \, \Omega_s^l  
\left [ f_0 ( \varepsilon_s  ) 
+  \varepsilon^{ (m) } \,  
f'_0 (\varepsilon_{ s} ) + \frac{1}{2} \left(  \varepsilon^{(m)} \right)^2  \,
  f^{\prime \prime}_0 ( \varepsilon_s  )   +  \frac{1}{6} \, 
\left(  \varepsilon^{(m)} \right)^3  \,
f_0^{ \prime \prime \prime} (\varepsilon_s ) + \order{ |\mathbf B|^4}  \right ]
\\ & \text{and }
\left( \sigma_s^{\mathrm{LF}} \right)_{ij} = -e^2 \tau \int \frac{d^3\mathbf{k}}{(2\pi)^3} \left[ \left(v_s\right)_i + \left(u_s\right)_i \right] 
f_0^\prime( \mathcal{E}_s) \frac{\partial \mathcal{Y}_s}{\partial E_j}\,,
\label{LF_cond}
\end{align}
respectively, where
\begin{align}
\label{LF_cond1}
{\boldsymbol u}_s= e  \, 
 (  {\boldsymbol  v}_s \cdot \mathbf{\Omega }_s  ) \,  {\mathbf B} \,, \quad
 \mathcal{Y}_s = \sum_{n=1}^{\infty} \left(e \,\tau\, 
D_s\right)^n \check{L}^n \left[  D_s \left\{ {\boldsymbol  v}_s +{\boldsymbol  u}_s \right\} \cdot \mathbf{E} \right],
\text{ and } \check{L} = ({\boldsymbol  v}_s \times \mathbf{B}) \cdot \nabla_{\mathbf{k}} \,.
\end{align}
Clearly, the first term in $ (\sigma^{\text{AH}}_s)_{ij} $ is independent of $\mathbf B$ (which is the origin of the nomenclature of ``anomalous Hall''), and it vanishes identically in our system. The nonzero terms therein appear only when we correctly account for the OMM-part, thus showing the importance of not omitting the OMM-contributions. The symbol $\check{L}$ denotes the so-calledLorentz-force operator and, comprising differentiation operator (with respect to $ \mathbf k$), it acts on everything appearing on its right-hand side. The detailed derivation of the expressions shown in Eqs.~\eqref{LF_cond}
and \eqref{LF_cond1} can be found in Refs.~\cite{ips_tilted_dirac, ips-spin1-ph}.
\end{enumerate}
We would like to point out that, while $\sigma^{\text{AH}}_s$ exclusively comprises terms which have odd powers of $|\mathbf B|$, $\sigma_s^{\text{LF}}$ contains even as well as odd powers of $ |\mathbf B|$. On the other hand, $\bar \sigma^s $ exclusively comprises only even powers of $|\mathbf B|$, which is expected by invoking the Onsager-Casimir reciprocity relations \cite{onsager1, onsager2, onsager3} --- therefore, retaining terms upto quadratic-in-$|\mathbf B|$ for $\bar \sigma^s $ ensures that it is correct upto $\order{ |\mathbf B|^3}$ overall. Consequently, all our final answers are correct upto $\order{ |\mathbf B|^3}$.

In the following, we will assume that a positive chemical potential $\mu$ is applied such $\mu>\Delta $, we will do all the calculations for the conduction band (i.e., we set $s = 2$), and we will employ the coordinate transformations shown in Eq.~\eqref{eqtrs} to perform the integrations. Henceforth, we will drop the band-index ($s$) in all the conductivity components. First, we divide up $\bar \sigma_{ij}$ as
\begin{align}
\bar \sigma_{ij} = \sigma_{ij}^{(d)} + \sigma_{ij}^{(bc)} + \sigma_{ij}^{(m)}\,,
\end{align}
where the superscripts of ``$(d)$'', ``$(bc)$'', and ``$(m)$'' are used to denote the Drude, BC-only, and the OMM-contributed parts, respectively. The Drude part is the one which is independent of the applied magnetic field, and is nonzero only for the longitudinal components [i.e., $ \sigma_{ij}^{(d)} \propto \delta_{ij} $]. The BC-only part does not contain any contribution from the OMM and, therefore, survives even when OMM is not included. The OMM-part is the one which goes to zero if we fail to include the OMM-induced corrections to the dispersion [i.e., $ \sigma_{ij}^{(m)} 
\vert_{{\boldsymbol{m}} \rightarrow \mathbf 0} = 0 $]. In a similar spirit, we use the notations,
\begin{align}
\label{LF_div}
\sigma^{\text{LF}}_{ij} = \sigma^{\text{LF},\text{H}}_{ij} +  \sigma^{\text{LF},bc}_{ij} + \sigma^{\text{LF},m}_{ij} \,,
\end{align}
to divide up the terms in the $\check{L}$-induced parts (of band $s=2$) according to their origins, which are all $ \mathbf B$-dependent. We would like to point out that the $\sigma^{\text{LF},\text{H}}_{ij}$-part is independent of the topological properties like BC and OMM (and this one includes the part giving rise to the conventional Hall effect).

Gathering all the nonzero terms from $I^{nz}_{1ij}$ [cf. Eq.~\eqref{i13}], we find that
$ \sigma_{ij}^{(bc)}$ is sourced by $  e^2  \, v^{(0, s)}_i \, v^{(0, s)}_j  \left( \mathbf  B \cdot \mathbf \Omega_s  \right )^2  \delta  (\varepsilon_s - \mu) $ and $ \sigma_{ij}^{(m)} \equiv \sigma_{ij}^{(pm)} +\sigma_{ij}^{(conc)} $ is sourced by  two distinct parts as follows:
\begin{align}
\sigma_{ij}^{(pm)}  & = \tau \, e^2 
\int \frac{d^3   {\mathbf k}} {(2  \, \pi)^{3}}\, I^{m}_{ij} , \quad
\sigma_{ij}^{(conc)}  = \tau \, e^2 
\int \frac{d^3   {\mathbf k}} {(2  \, \pi)^{3}} \, I^{conc}_{ij},
\nn I^{m}_{ij} & = 
v_i^{(m)} \,  v_j^{(m)}  \,  \delta  (\varepsilon_s - \mu) 
+
\varepsilon^{(m)}\left \lbrace v^{(0,s)}_j \, v_i^{(m)} + v^{(0, s)}_i  \, v_j^{(m)} \right \rbrace
\delta^\prime (\varepsilon_s - \mu) 
+ \frac { v^{(0, s)}_i \,v^{(0, s)}_j \left( \varepsilon^{(m)} \right)^2 
 \delta^{\prime \prime } (\varepsilon_s - \mu)}  {2}
\text{ and}\nn
I^{conc}_{ij}  & = -\, e  \left( \mathbf  B \cdot \mathbf \Omega_s  \right ) 
\left[
\left( v^{(0, s)}_i \,v_j^{(m)} + v_i^{(m)} \, v^{(0, s)}_j \right)   \delta  (\varepsilon_s - \mu)
+ \varepsilon^{(m)} \, v^{(0, s)}_i \,  v^{(0, s)}_j  
 \delta^\prime (\varepsilon_s - \mu)  \right ] .
\end{align}
Clearly, these two parts characterize the OMM-only and the concurrent BC-OMM contributions, respectively. This division will help us identify which part dominates and how the overall response is dictated by the dominant part.

The ranges of the values of the parameters in some realistic scenarios have been shown in Table
\ref{tab-params}, which we have used in our representative plots.
\begin{table}[h!]
\centering
\begin{tabular}{|c|c|c|}
\hline
 Parameter &     Natural Units  \\ \hline
$v_0$ from Ref.~\cite{phe_nlsm} & $0.0004$\\  
\hline
$v_z$ from Ref.~\cite{phe_nlsm} &  $0.00045$
\\  
\hline
$k_0$ from Ref.~\cite{phe_nlsm} &  $200 $  \text{eV} 
\\
\hline
$\Delta$ from Ref.~\cite{phe_nlsm} &  $0.02$  \text{eV}
\\ \hline
$\tau$ from Ref.~\cite{yang1} &  $15.2  $ eV$^{-1}$  
\\  \hline
$ B $ from Ref.~\cite{phe_nlsm} &  $ 0 $ --- $ 100 \, \text{eV$^{2}$} $ \\ \hline
$\mu$ from Ref.~\cite{phe_nlsm} &  $0.04 $ eV \\ \hline
\end{tabular}
\caption{\label{tab-params}The ranges of values for the various parameters, chosen to be used in the plots of conductivity, are tabulated here. While using the natural units, we need to set $\hbar=c=k_{B}=1$. The parameter values have been taken from \cite{phe_nlsm,yang1}.}
\end{table}
\\In the following, we will use the new variable,
\begin{align}
 \zeta=\sqrt{k_0^2 \,v_0^2 +\Delta^2 - \mu^2}\,. 
\end{align} 
 
\subsection{Set-up I: $ \mathbf{E} = E_x\, {\mathbf{\hat x}}$ and $
\mathbf{B} = B_x\,{\mathbf{\hat x}} + B_y\,{\mathbf{\hat y}}$}
\label{secset1}

In the set-up I shown in Fig.~\ref{figsetup}(b), we have $\mathbf{E} = E_x\, {\mathbf{\hat x}}$ and $
\mathbf{B} = B_x\,{\mathbf{\hat x}} + B_y\,{\mathbf{\hat y}}$.
Consequently, Eq.~\eqref{eqtopo} translates into $ \varepsilon^{(m)} (\mathbf{k}) 
 = \frac{e\,v_z\, v_0\, \Delta}
{2\, \epsilon^2}\, 
 \frac{B_x \,k_y - B_y\,k_x} {k_{\perp}},$ and
\begin{align}
\label{eqomm1}
v^{(m)}_x & =
 \frac{ - \, e\,  v_0\, v_z \, \Delta}{2\, \epsilon^4\, k_{\perp}^3} 
\left[ 2\, v_{0}^2\, k_x \left(k_{\perp}^{2} - k_{0}\, k_{\perp} \right)
\left( B_x\, k_y - B_y\, k_x \right) + \epsilon^2\, k_y
\left(B_x\, k_x + B_y\, k_y \right) \right]
\nn & = 
\frac{e \,  v_0\, v_z \, \Delta}{2 \, \epsilon^4} 
\left[
2 \, v_0^2 \, \kappa \cos{\phi}
\cos{\Phi} 
\left(   B_y \cos {\Phi} - B_x  \sin{\Phi} \right) 
- \frac{ \epsilon^2 \sin{\Phi}
\left( B_x \cos{\Phi}  + B_y \sin {\Phi} \right) }
{k_0 + \kappa \cos \phi}  
\right],
 \nn  v^{(m)}_y & = 
 \frac{ -\, e \, v_0\, v_z \, \Delta}{2\, \epsilon^4\, k_\perp^3} 
\left[ 2\, v_0^2\, k_y \left(k_\perp^{2} - k_0\,k_\perp\right)\, 
\left( B_x\, k_y - B_y\, k_x \right) - \epsilon^2\, k_x
 \left( B_x\, k_x + B_y\, k_y \right) 
\right]\nn
& = 
\frac{ e \, v_0\, v_z \, \Delta}{2 \, \epsilon^4} 
\left[ 2 \,v_0^2 \, \kappa \cos \phi 
\sin {\Phi}
\left(  B_y \cos{\Phi}  -  B_x \sin {\Phi} \right) 
+ \frac{ 
\epsilon^2 \cos{\Phi}
\left( B_y  \sin{\Phi} + B_x \cos {\Phi} \right) }
{k_0 + \kappa \cos \phi}  
\right],
 \nn v^{(m)}_z  & = 
\frac{ e \, v_0\, v_z^3 \, \Delta \,k_z }
{\epsilon^4\, k_\perp} 
\left( B_y\,k_x - B_x\,k_y \right)
=
\frac{e  \, v_0^2 \, v_z^2 \, \Delta \, \kappa \sin \phi } {\epsilon^4} 
\left(
B_y \, \cos{\Phi} - B_x  \sin{\Phi} \right ).
\end{align}
Plugging in these expressions in Eq.~\eqref{i1}, we arrive at
\begin{align}
\sigma^{(d)}_{xx} & = \frac{\tau \, e^2 \, v_0 \, k_0}
{8 \, \pi \, v_z \, \mu}  
\left(\mu^2 - \Delta^2 \right), \quad
\sigma^{(bc)}_{xx}  =
\frac{\tau \, e^4 \, v_z \, v_0^3  \, \Delta^2 
  \left( B_x^2 + 3 \, B_y^2 \right)}
{128 \, \pi \, \mu^7} 
 \, k_0 \left( \mu^2 - \Delta^2 \right) ,\nn
\sigma^{(m)}_{xx} & = 
\frac{\tau \, e^4 \,v_z \, v_0^3 \, \Delta^2 
\left( B_x^2 + 3 \, B_y^2 \right) }
{ 128 \, \pi \, \mu^7}
   \left [ 3\, k_0 \left( \Delta^2 - 2\, \mu^2 \right)
+ \frac{2 \,\mu^4}
{v_0\, \zeta} \right ],\nn
\sigma^{(d)}_{yx} & = 0 \,,
\quad
\sigma^{(bc)}_{yx}  =
\frac{ -\, \tau \, e^4 \, v_z \, v_0^3  \, \Delta^2 \, B_x \, B_y}
{ 64 \, \pi \, \mu^7} 
\, k_0 \left(  \mu^2 - \Delta^2 \right) ,\quad
\sigma^{(m)}_{yx}   = 
\frac{  -\, \tau \, e^4 \,v_z \, v_0^3 \, \Delta^2\,  B_x \, B_y }
{64 \, \pi \, \mu^7} 
 \left [ 3\, k_0\,\left( \Delta^2 
- 2\, \mu^2 \right)
+ \frac{2 \,\mu^4} 
{v_0\, \zeta} \right ]  ,
\nn \sigma^{\text{AH}}_{zx} & = 
\frac{ -\, e^3\, v_z \, v_0 \, k_0 \,\Delta^2 \, B_y}
 {16 \, \pi\, \mu^4} 
 \left (
1 +\frac{ 9\, e^2  \,v_z^2 \, v_0^2  \, \Delta^2 }
 { 4\, \mu^6 } \, {\mathbf B}^2  \right ) .
\end{align}
In particular, $ \sigma^{(pm)}_{xx}   =
\frac{ - \,\tau \, e^4 \,v_z \, v_0^3 \, \Delta^2 
\left( B_x^2 + 3 \, B_y^2 \right) }
{ 128 \, \pi \, \mu^7}
\left [k_0 \left(  6\, \mu^2 - 5\,\Delta^2  \right)
- \frac{2 \,\mu^4}
{v_0\, \zeta} \right ] $,
$  \sigma^{(conc)}_{xx}   =
\frac{-\,\tau \, e^4 \,v_z \, v_0^3 \, \Delta^4\,k_0 
\left( B_x^2 + 3 \, B_y^2 \right) }
{ 64 \, \pi \, \mu^7} $, 
$ \sigma^{(pm)}_{yx}    = 
\frac{  \tau \, e^4 \,v_z \, v_0^3 \, \Delta^2\,  B_x \, B_y }
{64 \, \pi \, \mu^7} 
 \left [k_0\,\left(  6\, \mu^2 - 5\,\Delta^2\right) - \frac{2 \,\mu^4} 
{v_0\, \zeta} \right ]$, and $ \sigma^{(conc)}_{yx}   = 
\frac{\tau \, e^4 \,v_z \, v_0^3 \, \Delta^4\,k_0\,  B_x \, B_y }
{32 \, \pi \, \mu^7}$. Our regime of validity is satisfied by $\lbrace \Delta ,\, \mu \rbrace \ll k_0$, because a toric Fermi surface is maintained for $\mathbf B = \mathbf 0 $ for these ranges. Hence, the terms $\propto k_0$ will dominate when we sum up $\sigma^{(bc)}_{ij}$, $\sigma^{(pm)}_{ij}$, and $\sigma^{(conc)}_{ij}$. We also note that the entire anomalous-Hall part of the conductivity is caused by the BC and the OMM concurrently.

\begin{figure}[t!]
\centering 
\subfigure{\includegraphics[width= 0.32 \textwidth]{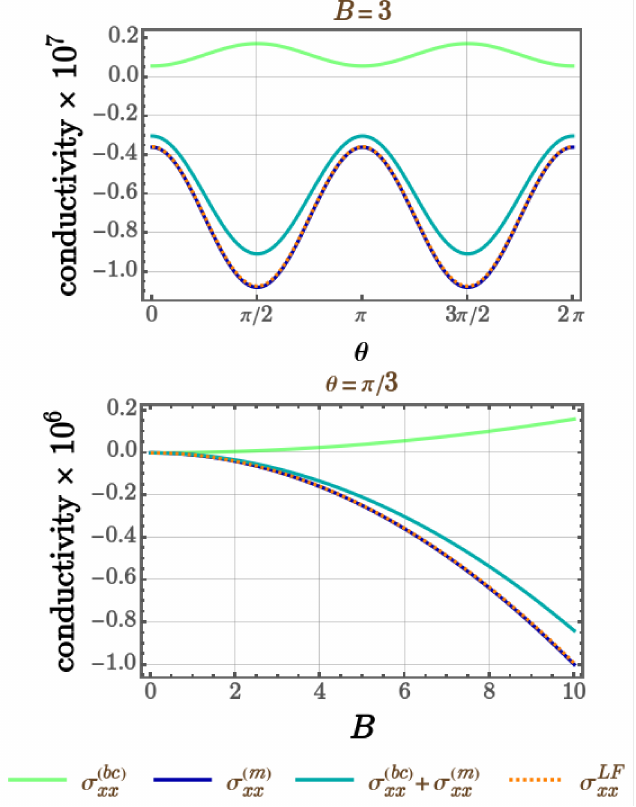}} \vrule
\subfigure{\includegraphics[width= 0.32 \textwidth]{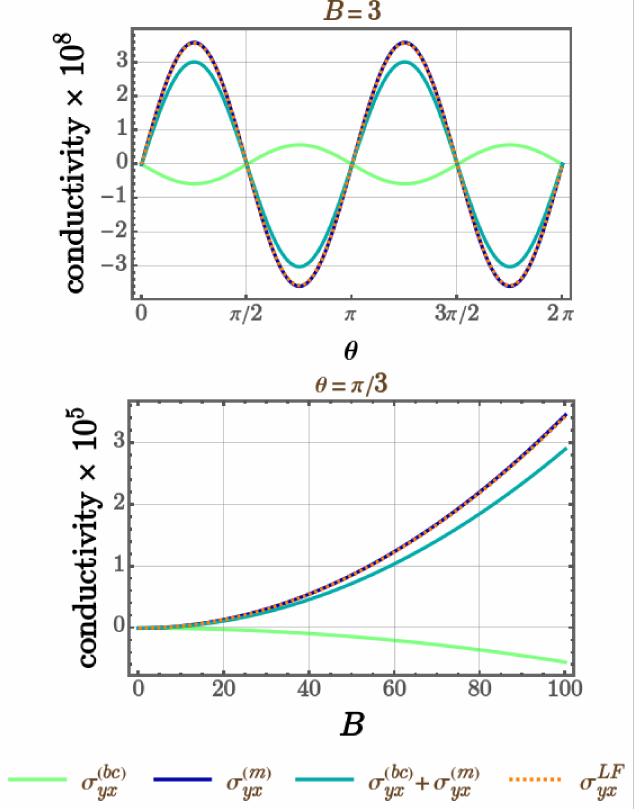}} \vrule \quad
\subfigure{\includegraphics[width= 0.305 \textwidth]{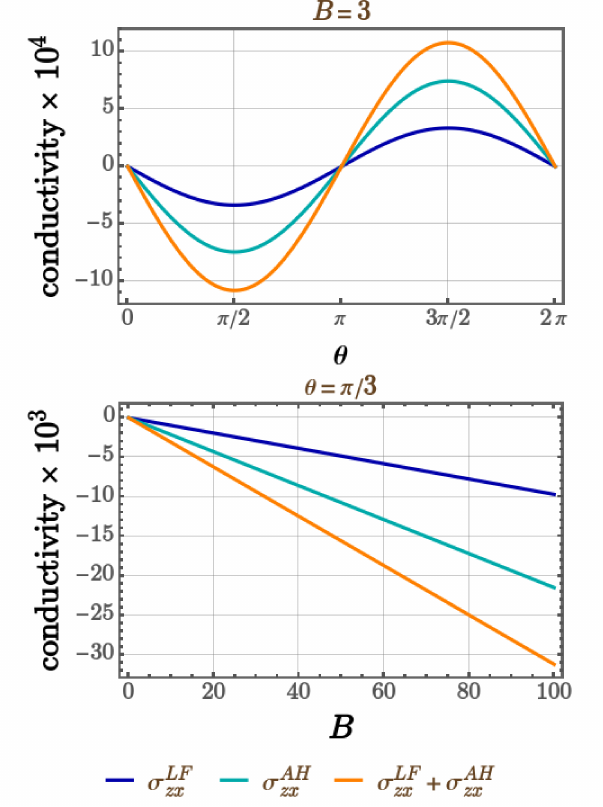}}
\caption{Set-up I: Behaviour of the three components of the conductivity tensor (in units of eV), taking into account all the contributions. We have used the parameter values from Table~\ref{tab-params}. The values of $B \equiv |\mathbf B|$ are in units of eV$^2$.
\label{figset1}}
\end{figure}

For obtaining the $\check L$-induced parts using Eq.~\eqref{LF_cond}, we outline below the various terms obtained by expanding the summation therein upto $n=3$:
\begin{enumerate}
\item $n = 1$ --- This part gives us nonzero values for the out-of-plane conductivity, which comprise
\begin{align}
\sigma^{\text{LF,H}}_{zx} & = \frac{ - \,\tau^{2} \, e^3 \, v_0 \,v_z\,k_0\,B_y}
{8 \, \pi \, \mu^{2}}  
\left(\mu^2 - \Delta^2 \right),\quad
\sigma^{\text{LF},bc}_{zx}  =
\frac{ -\, 9\,\tau^2 \, e^5 \,  v_z^3   \,v_0^3 \, \Delta^2 
  B_y\left( B_x^2 +  \, B_y^2 \right)}
{128 \, \pi \, \mu^8}  \, k_0 \left(\mu^2 - \Delta^2\right) ,\nn
\sigma^{\text{LF},m}_{zx} &  = 
\frac{  -\,3\, \tau^2 \, e^5 \,v_z^3 \, v_0^3 \, \Delta^2\,  B_y\left( B_x^2 +  \, B_y^2 \right) }
{128 \, \pi \, \mu^8} 
 \left [ k_0 \left( 7\,\Delta^2  - 10\, \mu^2 \right)
+ \frac{2 \,\mu^4}  {v_0\, \zeta} \right ] .
\end{align}
\item $n = 2$ --- For this part, we find that the nonzero components appear in the form of in-plane response, which are given by
\begin{align}
\sigma^{\text{LF},bc}_{xx} & =  \sigma^{\text{LF},m}_{xx}  = 0\,,
\quad \sigma^{\text{LF,H}}_{xx} =- \,
\frac{\tau^3 \, e^4\,  v_z \,v_0^4\,k_0^2 \left( B_x^2 + 3 \, B_y^2 \right) }
{ 16 \, \pi \, \mu^3}
   \left ( k_0\,v_0 - \zeta \right ), \nn
\sigma^{\text{LF},bc}_{yx} & =  \sigma^{\text{LF},m}_{yx}  = 0\,,
\quad\sigma^{\text{LF,H}}_{yx} = \frac{\tau^3 \, e^4 \,v_z \, v_0^4\,k_0^2\,  B_x \, B_y  }
{ 8 \, \pi \, \mu^3}
  \left ( k_0\,v_0 - \zeta \right ).  
\end{align}
\item $n = 3$ --- For this part, we find that the sole nonzero component is the non-BC non-OMM $zx$-component, captured by
\begin{align}
\sigma^{\text{LF,H}}_{zx} = \frac{3\, \tau^4 \, e^5 \,v_z^3 \, v_0^4\,k_0^2\,B_y \left( B_x^2 + B_y^2 \right) }
{ 16 \, \pi \, \mu^4}
   \left ( k_0\,v_0 - \zeta \right ).  
\end{align}
\end{enumerate}
Adding up all the out-of-plane contributions from $n=1$ and $ n=3$, we get
\begin{align}
\sigma^{\text{LF}}_{zx} & =\frac{ \tau^2 \, e^3 \,v_z \, v_0 \, B_y }
{128 \, \pi \, \mu^8} \Bigg[ 
3\,  e^2\,v_z^2\, v_0\left( B_x^2 +  \, B_y^2 \right) \Bigg\{  \Delta^2 \left(- \frac{2 \,\mu^4} 
{\zeta}+ k_0\,v_0 \left( 7\, \mu^2 -4\, \Delta^2 
 \right) \right) 
 + 8\, \tau^2\, v_0^2 \, k_0^2  \,\mu^4  \left(k_0\,v_0 - \zeta \right) \Bigg\} 
\nn & \hspace{ 2.85 cm } +  16\,k_0\,\mu^6(\Delta^2 - \mu^2) \Bigg].
\end{align}
Some representative curves are shown in Fig.~\ref{figset1}, where of course we have $\mu> \Delta$ (cf. Table~\ref{tab-params}).

\subsection{Set-up II: $\mathbf{E} = E_x \,{\mathbf{\hat x}}$ and $
\mathbf{B} = B_x\,{\mathbf{\hat x}} + B_z \,{\mathbf{\hat z}} $}
\label{secset2}

In the set-up II shown in Fig.~\ref{figsetup}(b), we have $\mathbf{E}  = E_x \,{\mathbf{\hat x}}$ and $
\mathbf{B} = B_x\,{\mathbf{\hat x}} + B_z \,{\mathbf{\hat z}}$.
Consequently, Eq.~\eqref{eqtopo} leads to $ \varepsilon^{(m)} (\mathbf{k}) = 
\frac{e\, v_z \,v_0 \, \Delta} {2\, \epsilon^2} \, \frac{B_x \,k_y} {k_{\perp}} ,$ and
\begin{align}
\label{eqomm2}
  v^{(m)}_x & = 
\frac{ -\, e\, \, v_z \,v_0\, \Delta \, k_x \,k_y \, B_x}
{2\, \epsilon^4 \,k_\perp^3}
\left[ 2\, v_0^2 \,\left( k_\perp^{2} - k_0\, k_\perp \right) 
+ \epsilon^2 \right]
= 
 \frac{ -\, e \, v_z \, v_0 \,\Delta \, B_x  \sin (2 \Phi)} 
 { 4\,\epsilon^4} 
\left( 2\, v_0^2\,  \kappa\cos{\phi} + \frac{ \epsilon^2  }
 { k_0 + \kappa \cos \phi} \right ), \nn
 v^{(m)}_y &= \frac{ - \, e \, v_z \,v_0 \, \Delta \,B_x }
 {2\, \epsilon^4\, k_\perp^{3}} 
\left[ 2\, k_y^2 \,v_0^2
 \left( k_\perp^{2} - k_0\, k_\perp \right)
- k_x^2\, \epsilon^2 \right]
 = \frac{ -\, e\, v_z \,v_0 \, \Delta\, B_x}  
 {2\, \epsilon^4} 
\left( 2\, v_0^2\,  \kappa \cos{\phi}  \sin^2{\Phi} 
- \frac{ \epsilon^2\,\cos^2 \Phi }
{k_0 +  \kappa \cos{\phi}} \right),\nn
v^{(m)}_z  & = 
\frac{ -\,e\, v_z^3 \, v_0 \, \Delta \, k_y\, k_z \, B_x}
{\epsilon^4\, k_\perp}
= 
\frac{ -\, e \, v_z^2 \, v_0^2\, \Delta \, B_x \,
\kappa \sin{\phi} \sin{\Phi}}
{\epsilon^4} \,.
\end{align}
Plugging in these expressions in Eq.~\eqref{i1}, we obtain
\begin{align}
\sigma^{(d)}_{xx} & = \frac{\tau \, e^2 \, v_0 \, k_0}
{8 \, \pi \,v_z \,\mu} 
 \left(\mu^2 - \Delta^2 \right),\quad
\sigma^{(bc)}_{xx}  =
\frac{\tau \, e^4  \,v_z\, v_0^3 \, \Delta^2 \, B_x^2}
{128 \, \pi \, \mu^7} \, \, k_0
\left( \mu^2 - \Delta^2 \right) , \nn
\sigma^{(m)}_{xx} & = 
\frac{ \tau\, e^4\, v_z \, v_0^3 \, \Delta^2\,B_x^2}
{128\, \pi\, \mu^7}
 \left[  3\, k_0 \left( \Delta^2 - 2 \, \mu^2 \right)
+ \frac{2 \,\mu^4}
{v_0\, \zeta} 
\right], \quad
\bar \sigma_{zx} =  (\sigma^{\text{AH}})_{yx}  =0\,.
\end{align}
Clearly, due to the special planar-vortex structure of $\mathbf \Omega_s (\mathbf k)$ and ${\boldsymbol{m}} ({ \mathbf k})$ (with a vanishing component along the $z$-direction), they do not couple with $ B_z \,{\mathbf{\hat z}} $. This is reflected by the absence of $B_z$ in any of the above expressions.
We also note that the in-plane transverse and the out-of-plane transverse components vanish.
Analyzing more deeply, we observe that $\sigma^{(pm)}_{xx}  = 
\frac{ \tau\, e^4\, v_z \, v_0^3 \, \Delta^2\,B_x^2}
{128\, \pi\, \mu^7}\left[k_0 \left( 5\,\Delta^2 - 6 \, \mu^2 \right)
+ \frac{2 \,\mu^4} {v_0\, \zeta} \right] $ and
$ \sigma^{(conc)}_{xx}  = \frac{ -\, \tau\, e^4\, v_z \, v_0^3 \, \Delta^4 \,k_0\,B_x^2}
{64\, \pi\, \mu^7}$.

For the $\check L$-induced parts, the various terms from the first three terms of the sum are summarised below:
\begin{enumerate}
\item $n = 1 $ --- Only the out-of-plane conductivity is nonzero, given by
\begin{align}
\sigma^{\text{LF,H}}_{yx}  &=  \sigma^{\text{LF},bc}_{yx}  = 0,\quad
\sigma^{\text{LF},m}_{yx}   = 
\frac{  \tau^2 \, e^5 \,v_z \, v_0^4 \, \Delta^2\,B_x^2\,B_z}
{32 \, \pi \, \mu^6} 
 \left [ 2\,k_0\,v_0\left( \frac{k_0\,v_0}{\zeta}-1 \right) +\frac{2\,\mu^2}{\zeta} -\frac{\mu^2 (\mu^2 - \Delta^2)}{\zeta^3} \right ] .
\end{align}

\item $n =2 $ --- Only the longitudinal components survive, captured by
\begin{align}
\sigma^{\text{LF},bc}_{xx} &=  \sigma^{\text{LF},m}_{xx}  = 0 \,,\nn
 \sigma^{\text{LF,H}}_{xx} & =  \frac{ - \,\tau^3 \, e^4 \,v_0^4\,k_0 }
{ 16 \, \pi \, \mu^3} \left[ 
\frac{2\,B_z^2\,v_0}{v_z} \left \lbrace 
2\,k_0^2\,v_0^2\left( \frac{k_0\,v_0}{\zeta}-1 \right) -(\mu^2 - \Delta^2)\right  \rbrace 
+ v_z \,k_0\,B_x^2
  \left ( k_0\,v_0 - \zeta \right ) \right].
\end{align}

\begin{figure}[t!]
\centering 
\subfigure{\includegraphics[width= 0.32 \textwidth]{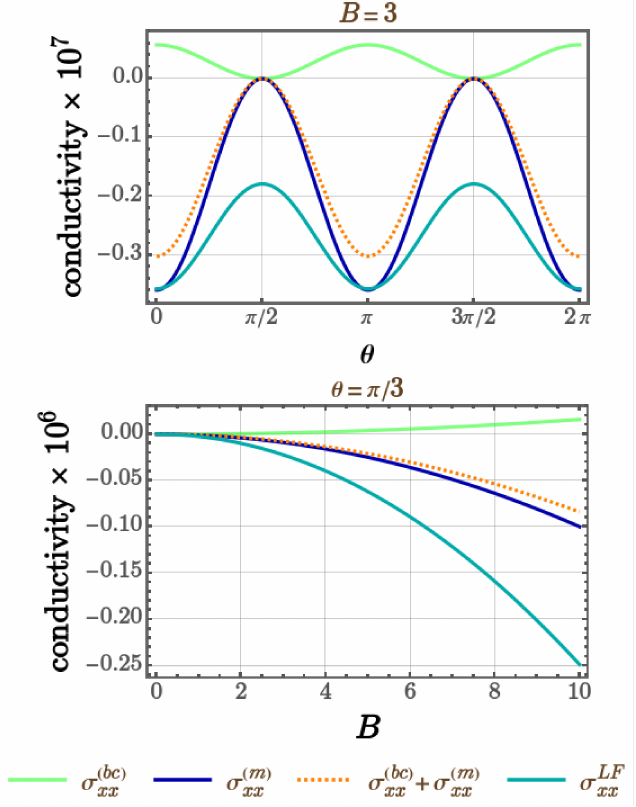}} 
\hspace{2 cm}\vrule \hspace{0.75 cm}
\subfigure{\includegraphics[width= 0.42 \textwidth]{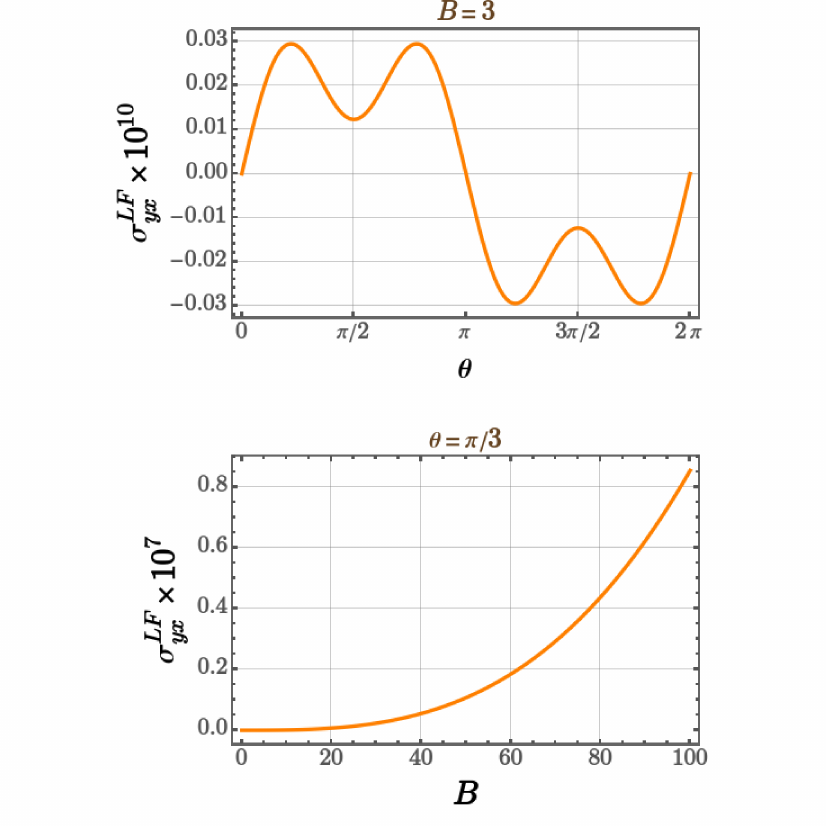}} 
\caption{Set-up II: Behaviour of the two nonzero components of the conductivity tensor (in units of eV), taking into account all the contributions. We have used the parameter values from Table~\ref{tab-params}. The values of $B \equiv |\mathbf B|$ are in units of eV$^2$.
\label{figset2}}
\end{figure}

\item $n = 3$ --- Only the out-of-plane component of the response survives, which is given by
\begin{align}
\sigma^{\text{LF,H}}_{yx} &= 
\frac{  \tau^4 \, e^5 \, v_z\, v_0^5\,k_0\,B_z}
{8 \, \pi \, \mu^4} 
\Bigg[ 
B_x^2 \left \lbrace 8\,k_0^2\,v_0^2\left( \frac{k_0\,v_0}{\zeta}-1 \right) 
-(\mu^2 - \Delta^2)\left( \frac{5\,k_0\,v_0}{\zeta}-1 \right)\right \rbrace \nn
& \hspace{ 3.5 cm }
- \,\frac{2\,v_0^2\,B_z^2}{v_z^2} \left \lbrace 4\,k_0^2\,v_0^2\left( \frac{k_0\,v_0}{\zeta}-1 \right) - (\mu^2 - \Delta^2)\left( \frac{\,k_0^3\,v_0^3}{\zeta^3}+1 \right)\right \rbrace \Bigg] .
\end{align}
\end{enumerate}
The overall out-of-plane contribution is obtained as
\begin{align}
\sigma^{\text{LF}}_{yx} & = \frac{ \tau^2 \, e^5 \,v_z \, v_0^4 \, B_z}  
{32\, \pi \, \mu^6}  \Bigg[ B_x^2\, \Delta^2 \left \lbrace 2 k_0\, v_0 \left( \frac{k_0\,v_0}{\zeta}-1 \right) + \frac{\mu^2 \left( 2\, k_0^2\,v_0^2 - 3 \left(\mu^2 -\Delta^2 \right) \right)}{\zeta^3}
 \right \rbrace \nn
& \hspace{ 3 cm  } 
+ 4\,\tau^2\, k_0\, v_0\, \mu^2 \Bigg\{
 \frac{2\, B_z^2\, v_0^2}{v_z^2}  \left(
4\, k_0^2\, v_0^2  \left( \frac{k_0\,v_0}{\zeta}+1 \right)
+ \mu^2 - \Delta^2 
- \frac{k_0^3 \,v_0^3 \left(8\, k_0^2\, v_0^2 -9 \left( \mu^2 - \Delta^2 \right) \right)}
{\zeta^3} \right) \nn & \hspace{ 5.5 cm}
 +   B_x^2 \left( \,8\, k_0^2\, v_0^2   \left( \frac{k_0\, v_0}{\zeta}-1\right) 
-  \left(\mu^2 - \Delta^2\right) \left( \frac{5\, k_0\, v_0}{\zeta}-1\right)\right)
\Bigg\} \Bigg].
\end{align}
Some representative curves are shown in Fig.~\ref{figset2}, where of course we have $\mu> \Delta$ (cf. Table~I).

\subsection{Set-up III: $\mathbf{E}= E_z\,{\mathbf{\hat z}}$ and
$\mathbf{B} = B_x\,{\mathbf{\hat x}} + B_z \,{\mathbf{\hat z}}$}
\label{secset3}

In the set-up III shown in Fig.~\ref{figsetup}(b), we have $\mathbf{E}= E_z\,{\mathbf{\hat z}}$ and $\mathbf{B} = B_x\,{\mathbf{\hat x}} + B_z \,{\mathbf{\hat z}}$.
Since the magnetic field is in the same plane as in set-up II, $\varepsilon^{(m)} (\mathbf{k}) $ and $ \boldsymbol
 v^{(m)} (\mathbf{k}) $ will be the same as in the previous subsection. Using those expressions in Eq.~\eqref{i1}, we obtain
\begin{align}
\sigma^{(d)}_{zz} & = \frac{\tau \, e^2 \, v_z \, k_0}
{ 4 \, \pi \, v_0 \,\mu } 
 \left(\mu^2 - \Delta^2 \right),\quad
\sigma^{(bc)}_{zz}  = 
\frac{\tau \, e^4  \,v_z^3\, v_0  \, \Delta^2\,B_x^2}
{ 32 \, \pi \, \mu^7} 
\, k_0 \left( \mu^2 - \Delta^2 \right) ,\quad
\sigma^{(m)}_{zz}  = 
\frac{ \tau\, e^4\, v_z^3\, v_0 \, \Delta^2\,B_x^2}
{32\, \pi\, \mu^7}
\, (-3 \,k_0) \left(   2\, \mu^2 - \Delta^2 \right),
\nn \bar \sigma_{xz}  & = 0 \,,\quad
(\sigma^{\text{AH}})_{yz}  = 
 \frac{ -\, e^3 \,v_z \,v_0\, k_0 \, \Delta^2 \, B_x} 
{ 16 \, \pi \,\mu^4 }
\left( 1
 + \frac{ 9\, e^2  \, v_z^2\, v_0^2 \, \Delta^2 }
 {  4 \, \mu^6 } \, B_x^2 \right ).
\end{align}
\begin{figure}[t!]
\centering 
\subfigure{\includegraphics[width= 0.33 \textwidth]{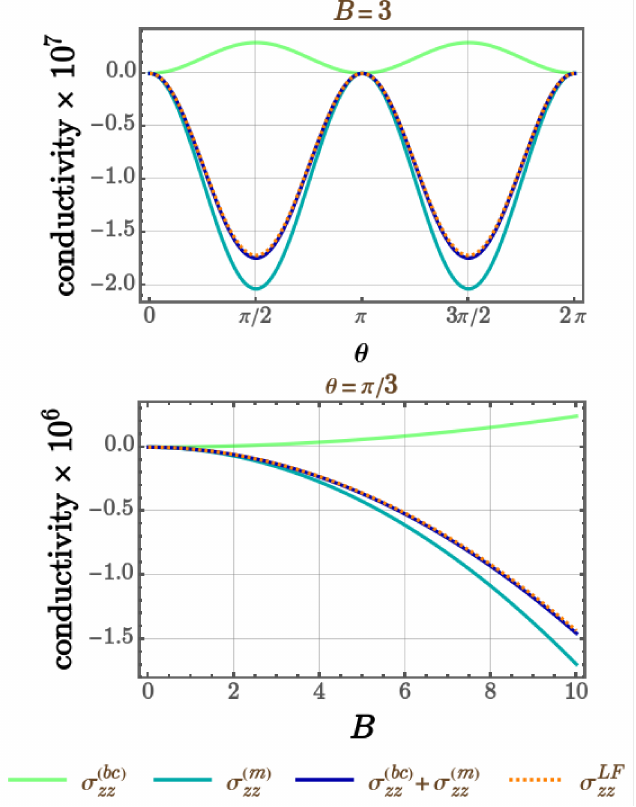}} 
\hspace{2 cm}\vrule \hspace{2 cm}
\subfigure{\includegraphics[width= 0.315 \textwidth]{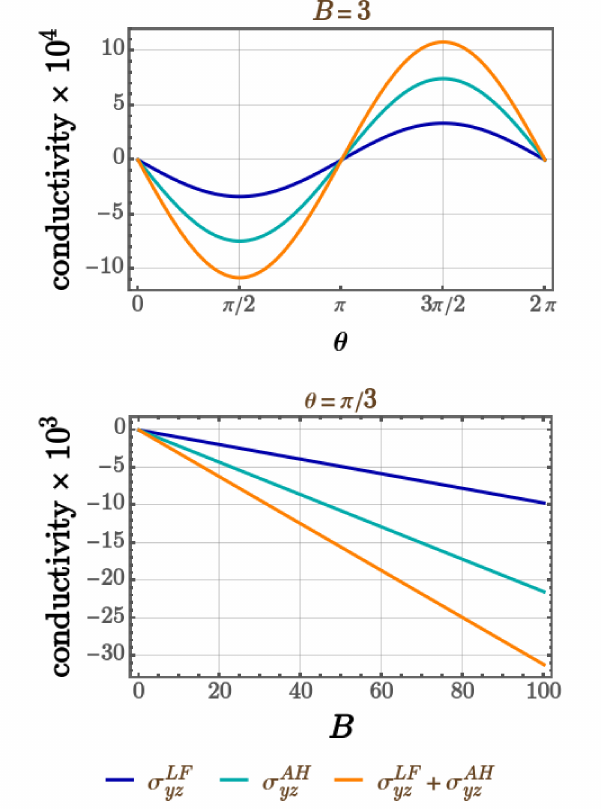}} 
\caption{Set-up III: Behaviour of the three components of the conductivity tensor (in units of eV), taking into account all the contributions. We have used the parameter values from Table~\ref{tab-params}. The values of $B \equiv |\mathbf B|$ are in units of eV$^2$.
\label{figset3}}
\end{figure}
Here, we observe that the $B_z$-component does not appear anywhere, which is again the artifact of $\Omega_s^z $ and $ m^z $ being zero. Furthermore, only the longitudinal (from $\bar \sigma_{i j}$) and out-of-plane components [from $(\sigma^{\text{AH}})_{ij}$] are nonzero. Dissecting deeper, we note that $ \sigma^{(pm)}_{zz}  = 
\frac{ - \, \tau\, e^4\, v_z^3\, v_0 \, \Delta^2\,B_x^2}
{32\, \pi\, \mu^7}
\, \,k_0 \left(   6\, \mu^2 -5\, \Delta^2 \right) $ and
$ \sigma^{(conc)}_{zz}  = 
\frac{ -\, \tau\, e^4\, v_z^3\, v_0 \, \Delta^2\, k_0\,B_x^2}
{16\, \pi\, \mu^7} $.

For the $\check L$-induced parts, the various terms from the first three terms of the sum are summarised below:
\begin{enumerate}
\item $n = 1$ --- The surviving ones are the out-of-plane components, which take the forms of 
\begin{align}
\sigma^{\text{LF,H}}_{yz} & = 
\frac{ -\,\tau^{2} \, e^3 \, v_z\,v_0 \,k_0\,B_x}
{8 \, \pi \, \mu^{2}}  
\left(\mu^2 - \Delta^2 \right),\quad
\sigma^{\text{LF},bc}_{yz}  =
\frac{  - \,9\,\tau^2 \, e^5 \, v_z^3 \, v_0^3  \, k_0\,\Delta^2 \,  B_x^3}
{128 \, \pi \, \mu^8} 
  \left(\mu^2 - \Delta^2 \right),\nn
\sigma^{\text{LF},m}_{yz} &  = 
\frac{   \tau^2 \, e^5 \,v_z \, v_0^3 \, \Delta^2\,B_x}
{128 \, \pi \, \mu^8} 
 \left[ 8\, v_0^2\,k_0\,B_z^2\, \mu^2\left( \frac{k_0\,v_0}{\zeta}-1 \right) 
 +  3\,v_z^2\,B_x^2\left\{k_0\, \left(10\,\mu^2-7\,\Delta^2\right)  -\frac{2\,\mu^4}{v_0\, \zeta} \right\} \right] .
\end{align}

\item $n =2 $ ---
The sole surviving part is a longitudinal component captured by
\begin{align}
\sigma^{\text{LF,H}}_{zz} = \frac{ - \,\tau^{3} \, e^4 \, v_0 \,v_z^3\,k_0\,B_x^2}
{8 \, \pi \, \mu^{3}}  
\left(\mu^2 - \Delta^2 \right).
\end{align}

\item $n = 3 $ ---
We find that an out-of-plane transverse component is the sole nonzero part, given by
\begin{align}
\sigma^{\text{LF,H}}_{yz} &= 
\frac{  \tau^4 \, e^5 \,  v_z\,v_0^4\,k_0\,B_x}
{16 \, \pi \, \mu^4} 
\left[ 
2\,v_0\,B_z^2 \left \lbrace 2\,k_0^2\,v_0^2
\left( \frac{k_0\,v_0}{\zeta}-1 \right) -(\mu^2 - \Delta^2)\right  \rbrace  + 3\,k_0\,v_z^2 \,B_x^2
  \left ( k_0\,v_0 - \zeta \right ) \right] .
\end{align}
\end{enumerate}
On adding up all the out-of-plane parts, we get
\begin{align}
\sigma^{\text{LF}}_{yz} &=
\frac{ \tau^2\, e^3\,v_z\, v_0\, B_x}{128\, \pi\, \mu^8} 
\Bigg[ 
e^2\, v_0\,\Delta^2
\Bigg\{ 3\, B_x^2\, v_z^2\,\left(  - \,\frac{2\, \mu^4}{\zeta} 
+  k_0\, v_0 \left( 7\, \mu^2 - 4\,\Delta^2 \right)
\right) + 
  8\, v_0^3\,k_0\, \mu^2\, B_z^2 
  \left( \frac{ k_0\, v_0}{\zeta} - 1 \right) \Bigg\}\nn  &
\hspace{ 2.85 cm} + 8 \,\tau^2\, e^2\, v_0^3\, k_0\,\mu^4
\Bigg\{ 2\, v_0\, B_z^2
\left( 2\,k_0^2\, v_0^2 \left( \frac{ k_0\, v_0}{\zeta} - 1 \right) 
-\left( \mu^2-\Delta^2  \right)\right) 
 + 3\, v_z^2\,k_0\,B_x^2 \left(k_0\, v_0- \zeta \right) \Bigg\}
\nn & \hspace{ 2.85 cm } -16\, k_0\, \mu^6 \left(\mu^2-\Delta^2  \right) \Bigg].
\end{align}
Some representative curves are shown in Fig.~\ref{figset3}, where of course we have $\mu> \Delta$ (cf. Table~I).

\subsection{Discussion and comparison of the results}

For the ease of the reader, we provide a summary of the results for the three set-ups in Tables~\ref{tab_all}(a)-\ref{tab_all}(d). The overall characteristics are also demonstrated by the curves of Figs.~\ref{figset1}-\ref{figset3}, choosing some representative parameter values from the literature (cf. Table \ref{tab-params}).

\begin{table}[h!]
\subfigure[Nonzero contributions for $\bar \sigma_{i j}$ and $ \sigma^{\text{AH}}_{ij} $ for all the three set-ups.]
{\centering
\begin{tabular}{|c|c|c|c|}
\hline
    & Longitudinal
    & In-plane transverse
    & Out-of-plane transverse \\ \hline
Set-up I
& $ \Upsilon_1 \left( B_x^2 +3\, B_y^2 \right) $ 
&  $ -\,2\, \Upsilon_1 \, B_x \, B_y$
&  $ \Upsilon_3 \, B_y
\left( 1 + \Upsilon_4 \, \mathbf B^2 \right) $ \\ \hline
Set-up II
& $ \Upsilon_1 B_x^2$
& 0
& 0  \\ \hline
Set-up III
& $ \Upsilon_2 \, B_x^2$
& 0
& $ \Upsilon_3 \, B_x
\left( 1 + \Upsilon_4 \, B_x^2 \right) $ \\ \hline
\end{tabular}}
\subfigure[Set-up I: Nonzero contributions for $ \sigma^{\text{LF}}_{i j}$.]
{\centering
\centering
\begin{tabular}{|c|c|c|c|c|c|}
\hline
 & Longitudinal ($\sigma_{xx}$) & In-plane transverse ($\sigma_{yx}$)  & \multicolumn{3}{c|}{Out-of-plane transverse ($\sigma_{zx}$)} \\ \hline
    & $\sigma^{\text{LF},\text{H}}_{xx}$ & $\sigma^{\text{LF},\text{H}}_{yx}$
    & $\sigma^{\text{LF},\text{H}}_{zx}$ & $\sigma^{\text{LF},bc}_{yx}$ & 
    $\sigma^{\text{LF},m}_{yx}$ \\ \hline
$n=1$ & $ 0$  & $0 $ & terms $\propto  B_y $ & terms $\propto  B_y \left( B_x^2 + B_y^2 \right) $
&  terms $\propto  B_y \left( B_x^2 + B_y^2 \right) $ \\ \hline
$n=2$ &terms $\propto  \left( B_x^2 +3\, B_y^2 \right) $ & terms $\propto   B_x \, B_y $ & 
$0$  & $0$  & $0$ \\ \hline
$ n=3 $ & $0$ & 0 & terms $\propto  B_y\left( B_x^2 + B_y^2 \right) $ & $0$ & $0$
\\ \hline
\end{tabular}}
\subfigure[Set-up II: Nonzero contributions for $  \sigma^{\text{LF}}_{i j}$.]
{\centering
\centering
\begin{tabular}{|c|c|c|c|c|}
\hline
 & Longitudinal ($\sigma_{xx}$)   & \multicolumn{2}{c|}{Out-of-plane transverse ($\sigma_{yx}$)} \\ \hline
& $\sigma^{\text{LF},\text{H}}_{xx}$  & $\sigma^{\text{LF},\text{H}}_{yx}$ &  $ \sigma^{\text{LF}, m}_{yx}$ \\ \hline
$n=1$ & $0$& $0$ & terms $\propto  B_z  \,B_x^2$  \\ \hline
$n=2$ & terms $\propto  B_z^2 $ and $B_x^2  $  & $0$  & $0$ \\ \hline
$ n=3 $ & $0$ & terms $\propto  B_z \, B_x^2$ and $ B_z^2 $ & $0$ 
\\ \hline
\end{tabular}}
\subfigure[Set-up III: Nonzero contributions for $  \sigma^{\text{LF}}_{i j}$.]
{\centering
\begin{tabular}{|c|c|c|c|c|}
\hline
& Longitudinal ($\sigma_{zz}$)  & \multicolumn{3}{c|}{Out-of-plane transverse ($\sigma_{yz}$)} \\ \hline
& $\sigma^{\text{LF},\text{H}}_{zz}$ & $\sigma^{\text{LF},\text{H}}_{yz}$ & $\sigma^{\text{LF},bc}_{yz}$ & 
    $\sigma^{\text{LF},m}_{yz}$ \\ \hline
$n=1$ & 0  &  terms $\propto B_x$  &  terms $\propto  B_x^3 $ & terms $\propto  B_x\, B_z^2  $ and $B_x^3$  \\ \hline
$n=2$ & terms $\propto  B_x^2  $  & $0$ & $0$  & $0$ \\ \hline
$ n=3 $ & $0$ & terms $\propto  B_x \, B_z^2 $ and $ B_x^3 $ & $0$ & $0$
\\ \hline
\end{tabular}}
\caption{\label{tab_all}Comparison of the overall behaviour of all nonzero components of the magnetoelectric conductivity for the three set-ups, when the chemical potential cuts the $s=2$ band at $T=0$.}
\end{table}

First, let us discuss the non-$\check L$-induced terms, considering the $\mathbf B$-dependent (i.e., non-Drude) parts. For set-up II, only the longitudinal components are nonzero. The observations of Table~\ref{tab_all}(a) are explained below:
\begin{enumerate}

\item The longitudinal components for set-ups I and II are proportional to $\left( B_x^2 +3\, B_y^2 \right)$ and $B_x^2$, respectively, with the same proportionality constant of
\begin{align} 
\Upsilon_1  =  \frac{\tau \, e^4 \, v_z \, v_0^3  \, \Delta^2 }
{128 \, \pi \, \mu^7} \left[
k_0 \left( \mu^2 - \Delta^2 \right) 
+
\left \lbrace - \, 3\, k_0 \left( 2 \, \mu^2 - \Delta^2 \right)
+ \frac{2 \,\mu^4}
{v_0\, \zeta}  
\right \rbrace \right ].
\end{align} 
The terms within the curly brackets represent the OMM-contributed part and, clearly, it is comparable to the BC-only part. Hence, it is quintessential to take the OMM-effects into account in order to capture the correct behaviour of the conductivity. The BC-only and OMM-parts appear with opposite signs and, hence, reduce the overall magnitude of the response.
In fact, since $\lbrace \Delta ,\, \mu \rbrace \ll k_0$ in the regime of our interest (when a torus shape of the Fermi surface is maintained for $\mathbf B = \mathbf 0 $), we find that the OMM-part dominates over the BC-only part --- this results in a change in sign of the overall response compared to the case when OMM is ignored (cf. leftmost panels of Figs.~\ref{figset1} and \ref{figset2}). 

\item For the longitudinal component of set-up III, it is seen to be proportional to $B_x^2$, with the proportionality constant of
\begin{align} 
\Upsilon_2  =  \frac{\tau \, e^4  \,v_z^3\, v_0 \,k_0 \, \Delta^2}
{ 32 \, \pi \, \mu^7} 
\left[  \left( \mu^2 - \Delta^2 \right)
-3\left(   2 \, \mu^2 - \Delta^2 \right)
\right ] .
\end{align}
The second term within the square brackets is the OMM-contributed part. Here too we observe that the BC-only and the OMM-induced parts come with opposite signs, with the magnitude of the latter dominating over the former (cf. left panel of Fig.~\ref{figset3}).

\item The in-plane transverse component for set-up I is given by $ (- \,2\, \Upsilon_1 )
\, B_x\, B_y  $, thus harbouring the same opposing effects of the BC-only and OMM parts (cf. middle panel of Fig.~\ref{figset1}). The in-plane transverse components for set-ups II and III are identically zero.

\item Table~\ref{tab_all}(a) also shows the out-of-plane components arising from the anomalous-Hall effect, which vanish for set-up II. For set-ups I and III, they take the forms of $ \Upsilon_3 \, B_y
\left( 1 + \Upsilon_4 \, \mathbf B^2 \right) $ and $ \Upsilon_3 \, B_x
\left( 1 + \Upsilon_4 \, B_x^2 \right) $, respectively, with
\begin{align}
\Upsilon_3  = \frac{ -\, e^3\, v_z \, v_0 \, k_0 \,\Delta^2 }
 {16 \, \pi\, \mu^4} \,,\quad
 \Upsilon_4  = \frac{ 9\, e^2  \,v_z^2 \, v_0^2  \, \Delta^2 }
 { 4\, \mu^6 }  \,.
\end{align}
Here, these terms solely arise from the OMM-contributed parts, thus emphasizing once more the importance of not neglecting the OMM corrections.

\end{enumerate}
Overall, the anisotropic response is captured by the distinct forms of the electric conductivity in the three distinct set-ups. Indeed, it is completely expected (even before doing an explicit calculation) that the $B_z$-components will not appear at all in the topology-induced nonzero response. This is due to the fact that the BC and the OMM have nonvanishing components only along the $x$- and $y$-directions, forming vortex-like flux lines. Consequently, it is no surprise that the topology-induced transverse components vanish altogether in the set-ups II and III.

Next come the $\check L$-induced terms, which are summarised in Tables~\ref{tab_all}(b)-\ref{tab_all}(d) for the three set-ups separately. The out-of-plane response is expected from the Lorentz force (corresponding to the conventional Hall effect). However, in addition, we observe here that the $\check L$-operator also gives rise to in-plane longitudinal and transverse currents, also noted in Refs.~\cite{ips_tilted_dirac, ips-spin1-ph}. Furthermore, there exist out-of-plane terms beyond the classical Hall term. Below, we elaborate on the notable characteristics:

\begin{enumerate}

\item Set-up I --- The curves of $\sigma^{\text{LF}}_{xx}$ and $\sigma^{\text{LF}}_{yx}$ almost coincide with those of $  \sigma^{(m)}_{xx} $ and $\sigma^{(m)}_{yx}$, respectively, as seen from the leftmost and middle panels of Fig.~\ref{figset1}. $\sigma^{\text{LF}}_{zx}$ has the same sign as $\sigma^{\text{AH}}_{zx}$, which thus reinforce each other by adding up, as reflected in the rightmost panel of Fig.~\ref{figset1}.

\item Set-up II --- $\sigma^{\text{LF}}_{xx}$ has a higher magnitude than $  (\sigma^{(bc)}_{xx} + \sigma^{(m)}_{xx})$, with each possessing an overall negative sign. This is reflected in the curves illustrated in the left panel of Fig.~\ref{figset2}. The out-of-plane component is solely contributed by $\sigma^{\text{LF}}_{yx}$, as $\sigma^{\text{AH}}_{yx}$ is identically zero.

\item Set-up III --- $\sigma^{\text{LF}}_{zz}$ almost coincides with $  (\sigma^{(bc)}_{zz} + \sigma^{(m)}_{zz} )$, as observed in the left panel of Fig.~\ref{figset3}. In analogy with set-up I, $\sigma^{\text{LF}}_{yx}$ has the same sign as $\sigma^{\text{AH}}_{yx}$, which reinforce each other by the addition of the overall magnitudes, and are demonstrated by the curves in the right panel of Fig.~\ref{figset3}.

\end{enumerate}
All the non-Drude and non-LF response-values originate purely from topological quantities, viz. the BC and the OMM, which in turn are proportional to a nonzero $\Delta $. Therefore, these vanish for an ungapped nodal-ring where the $\mathcal P \mathcal T$-symmetry is preserved. This property can be contrasted with the features of nodal-point semimetals, where non zero BC and OMM exist for ungapped systems, giving rise to chiral response in the form of in-plane conductivity \cite{ips_rahul_ph_strain, rahul-jpcm, ips-kush-review, ips-ruiz, ips-tilted, ips-rsw-ph, ips-shreya, ips-spin1-ph}. In particular, the signs of the conductivity there are not determined by a competition of $\Delta$ and $\mu$ because the systems are intrinsically gapless at the nodal points.

\section{Summary and outlook}
\label{secsum}

The detection of topological properties of 3d semimetallic bandstructures via linear response in planar-Hall set-ups has garnered tremendous attention in contemporary research, spanning both theoretical and experimental studies. In this paper, we contribute to such efforts by computing the magnetoelectric conductivity considering differing orientations of a gapped nodal-ring with respect to the $\mathbf E \mathbf B$-plane. Since we have considered the simple case of untilted NLSMs, the in-plane components comprise only even powers of $|\mathbf B |$. The appropriate inclusion of the OMM leads to nonzero out-of-plane components from the anomalous-Hall effects, which, otherwise, would not show up if the OMM was omitted. All our results show that the OMM must be considered at an equal footing with the BC, and that it cannot be ignored without the risk of missing important contributions to the net conductivity. A crucial aspect of our analysis is the inclusion of the terms induced by the recursive operation of the Lorentz-force operator (viz. $\check{L}$), which has been largely ignored (other than the part causing the conventional Hall effect) in the literature while investigating transport in semimetals. However, using the techniques that our group has developed in our earlier works \cite{ips_tilted_dirac, ips-spin1-ph}, we have demonstrated explicitly that the action of $ \check L $ gives rise to in-plane conductivity terms, in addition to the out-of-plane ones, which are comparable in magnitude with the terms contained in $\bar \sigma^s_{i j}$.

Our earlier works on planar-Hall set-ups involved the consideration of nodal-point semimetals, such as Weyl/multi-Weyl nodes \cite{ips_rahul_ph_strain, ips-kush-review, rahul-jpcm,ips-ruiz, ips-tilted}, Rarita-Schwinger-Weyl semimetals \cite{ips-rsw-ph, ips-shreya}, and triple-point semimetals~\cite{ips-spin1-ph}. In particular, we have studied the interplay of direction-dependence and topology in anisotropic systems like the multi-Weyl nodes. In contrast with their behaviour, the NLSMs have nonzero values of BC and OMM only in the presence of a finite mass-gap $\Delta $. Even with a nonzero $\Delta $, the BC and OMM have zero components in the direction perpendicular to the nodal-ring's plane. As a consequence, one or both of the transverse components of the conductivity vanish for particular choices of the orientation of the $\mathbf E \mathbf B$-plane. We note that this is not the case for multi-Weyl semimetals \cite{rahul-jpcm, ips-tilted}.

Here, we have only shown the results for the electrical conductivity. One could also derive the analogous expressions for the thermoelectric-conductivity tensor ($\alpha^s $) and magnetothermal coefficient ($\kappa^s $), repeating a similar exercise, but at a finite temperature \cite{ips_rahul_ph_strain, rahul-jpcm, ips-ruiz, ips-shreya, ips-spin1-ph}. However, we have not ventured into computing those, because the Mott relation and Wiedemann-Franz law have been shown to hold for all these set-ups \cite{xiao06_berry}, which allow us to easily infer the forms of the $\alpha^s_{ij} (T) $ and $\kappa^s_{ij} (T) $, once we know the expression of $\sigma^s_{ij} (T=0) $ [after using Eq.~\eqref{eqsigmat}].

In the future, it will be rewarding to repeat our calculations in the presence of nonzero tilts of the NLSMs, in the same spirit as we have done for tilted Weyl/multi-Weyl nodes \cite{rahul-jpcm, ips-tilted, ips-shreya, ips_tilted_dirac}. In particular, tilting will manifest itself by causing linear-in-$|\mathbf B| $ terms to materialize in the in-plane response coefficients~\cite{zyuzin_tilt_dirac, rahul-jpcm, ips-tilted, das-agarwal_omm, ips-tilted, ips-shreya, ips_tilted_dirac}.
Next, it will be worthwhile to study the transport properties under a strong quantizing magnetic field, when it is quintessential to incorporate the quantization of the dispersion into discrete Landau levels \cite{ips-kush, fu22_thermoelectric, staalhammar20_magneto, yadav23_magneto}. Yet another interesting avenue is to consider a non-flat (in energy) nodal-ring, which might give rise to nontrivial scatterings between concyclic points, analogous to internode scatterings in nodal-point semimetals~\cite{ips-internode}. 
Last but not the least, if we wish to quantitatively explore realistic scenarios, the effects of disorder and/or many-body interactions invariably come into play. To analyze correlated physics, we have to employ state-of-the-art many-body techniques (such as Green's functions) to compute the resulting response \cite{ips-seb, ips_cpge,  ips-klaus, rahul-sid, ipsita-rahul-qbt, ips-qbt-sc, ips-biref, ips-hermann-review}.

\bibliography{ref_nl}

\end{document}